%% file: main.tex
\documentclass[]{spie}  

\usepackage{amsmath,amsfonts,amssymb}
\usepackage{graphicx}
\usepackage{xcolor}
\usepackage[colorlinks=true, allcolors=blue]{hyperref}
\usepackage{array}
\usepackage{multirow}
\usepackage{float}
\usepackage{eurosym}

\title{Guiding Design Choices for Wide-Field IFS: Trade-Offs Between Replication and Complexity for WST}

\author[a]{C. Cudennec}
\author[a]{A. Jeanneau}
\author[a]{R. Bacon}
\author[b]{T. Lépine}
\author[a]{M. Lehnert}
\author[a]{R. Giroud}
\author[a]{J-E. Migniau}
\author[c]{D. Lee}
\author[d]{R. de Jong}
\author[e]{L. Fréour}
\affil[a]{Université Claude Bernard Lyon 1, CNRS, Centre de Recherche Astrophysique de Lyon, UMR5574, Saint-Genis-Laval, France}
\affil[b]{Université Jean Monnet, CNRS, Institut d’Optique
Graduate School, Laboratoire Hubert Curien, UMR 5516, Saint-Etienne, France}
\affil[c]{STFC UK Astronomy Technology Centre, Royal Observatory Edinburgh, Blackford Hill, Edinburgh, EH9 3HJ, United Kingdom}
\affil[d]{Leibniz-Institut für Astrophysik Potsdam (AIP), Potsdam, Germany}
\affil[e]{Department of Astrophysics, University of Vienna, Türkenschanzstrasse 17, A-1180 Vienna, Austria}

\authorinfo{Send correspondence to corentin.cudennec@univ-lyon1.fr}

\begin{document} 
\maketitle

\input{abstract}


\input{1-Introduction}

\input{2-Methodology_and_key_parameters}

\input{3-Evaluation_metrics}

\input{4-Draft_optical_designs}

\input{5-Trade-off_matrix}

\input{6-Conlusion}

\appendix
\input{Appendix}

\acknowledgments 
 
This project has received funding from the European Union’s Horizon Europe research and innovation programme under grant agreement No 101183153.

\bibliography{report} 
\bibliographystyle{spiebib} 

\end{document}

%% file: abstract.tex
\begin{abstract}

The Wide-field Spectroscopic Telescope\footnote{\url{https://www.wstelescope.eu/}} (WST) is a proposed 12-meter segmented facility optimized for seeing- and Ground Layer Adaptive Optics-limited observations in the visible and designed to operate both a high-multiplex multi-object spectrograph and a panoramic integral field spectrograph (IFS). The WST IFS concept builds on instruments such as MUSE at the VLT (Very Large Telescope) \cite{MUSE}, using field splitters and image slicers to reformat a large field into pseudo-slits feeding spectrographs with two optimized spectral channels.

This paper presents the integrated design approach adopted for the IFS, focusing on a trade study of spectrograph architectures. We explore design choices such as pixel pitch, detector format, and camera optical design against throughput, image quality, error budgets, volume, cost. The study adds one ecological metric: the carbon footprint of building each spectrograph, to inform design sustainability. The study also explores the potential of curved detectors. Early results suggest that many simpler spectrographs outperform fewer complex units technically and economically.

\end{abstract}

%% file: 1-Introduction.tex
\section{INTRODUCTION}

Over the next decade, the imaging capabilities of some of the new facilities (e.g., LSST/VRO, SKAO, CTA, JWST, Euclid, NGRST, Athena) will detect and classify a huge number of astronomical objects. However, to learn more about them, a spectroscopic follow-up is required. And given the expected number of sources, only a dedicated spectroscopic facility will be able to fully realize the scientific potential of these wide-field imaging surveys~\cite{WSTWP}. A large collaboration of 23 research institutions and universities across 10 nations is currently working on developing such a facility, aiming to propose it to the European Southern Observatory (ESO) as its next innovative ground-based programme via via the Expanding Horizons programme\footnote{\url{https://next.eso.org/}}.

\begin{figure} [H]
\begin{center}
\begin{tabular}{c} 
\includegraphics[height=9.5cm]{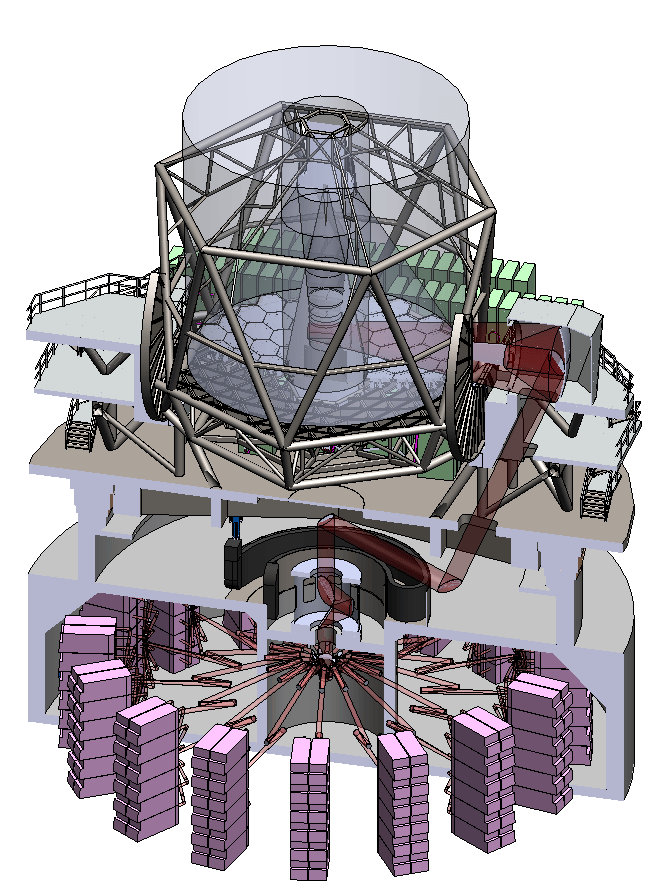}
\end{tabular}
\end{center}
\caption[SC] 
{ \label{fig:WST} 
Cross-section of the facility’s current layout. The IFS is located in the bottom part of the facility; showing 192 IFUs (pink boxes) and the associated relays (red beams).}
\end{figure}

WST is designed to operate two complementary instruments simultaneously (see Fig.~\ref{fig:WST}): a high-multiplex multi-object spectrograph and a panoramic integral field spectrograph (IFS). The IFS represents a major step forward compared to existing similar instruments, thanks to its spectral coverage, field of view and collecting area. To be precise, it will cover 370 -- 930 nm, observe a field of view of 3x3 arcmin² --- 9 times as large as MUSE or BlueMUSE --- and with an effective collecting area twice that of the VLT.

Realising an instrument of this scale requires confronting a set of architectural decisions that have no clear precedent in the IFS literature, even though there was already support favouring replicated architectures for high-étendue instruments\cite{Gary}. The spectrographs will constitute a dominant fraction of the overall facility in terms of volume, cost, and integration complexity, making a rigorous trade-off study not merely useful but essential. At the heart of this study lies a central tension: \textbf{should one build a large number of relatively simple spectrographs}, exploiting economies of scale as pioneered in optical astronomy by VIRUS\cite{VIRUS_intro} and MUSE\cite{MUSE}, \textbf{or a smaller number of individually more complex units?}

The architecture of an IFS is mostly driven by its spectrographs; the upstream subsystems act as relays, and their characteristics can, for the most part, be adapted to what the spectrographs require. Hence, in this paper we explore a wide range of design choices including camera optical design, detector format and pixel pitch, and spatial sampling, and evaluate them against a comprehensive set of metrics: throughput, image quality, error budgets, optomechanical volume, cost, and carbon footprint. Although the question is posed in the context of WST, the framework and tools developed here (in particular the cost model and the environmental impact assessment) are intended to be broadly applicable to the design of any future wide-field IFS or multi-unit spectrograph facility.

The potential of curved detectors to simplify camera designs and potentially improve throughput and cost is examined throughout, and a preliminary risk assessment informed by manufacturer consultation is provided. Early results from both the analytical models and the draft optical designs consistently converge on the same conclusion: many simpler spectrographs outperform fewer complex units, technically and economically.

The paper is organised as follows: Sec.~\ref{sec: methodo} defines the parameter space and key design drivers, including étendue, anamorphism, pupil diameter scaling, and spectral resolution. Sec.~\ref{sec:metrics} presents the evaluation metrics used to filter and compare design options. Sec.~\ref{sec:designs} carries out the actual design trade-off of the filtered options, with draft optical designs, volume and cost comparisons, and an environmental impact assessment. Sec.~\ref{sec:tradeoff} synthesises these results in a weighted trade-off matrix. Sec.~\ref{sec:conclusions} concludes and identifies the most promising design families for the WST IFS. Supporting derivations and calibration data are provided in the appendices.

%% file: 2-Methodology_and_key_parameters.tex
\newpage
\section{METHODOLOGY AND KEY PARAMETERS}
\label{sec: methodo}

This section outlines the process and criteria used to evaluate and compare the different spectrograph design options. Given that the spectrographs will constitute a significant fraction of the overall facility in terms of both volume and cost, conducting as comprehensive a study as possible is essential. We begin by presenting the range of design choices considered, including camera designs, detector types, and spatial sampling. We then detail the methodology used to explore the architecture space and to filter design options to be drafted, and define comparison metrics such as cost, transmission and carbon footprint. Some sections, while not directly part of the trade-off analysis, provide important context, for instance the rationale behind estimating the pupil diameter prior to drafting any design.

\begin{figure} [H]
\begin{center}
\begin{tabular}{c} 
\includegraphics[height=5.4cm]{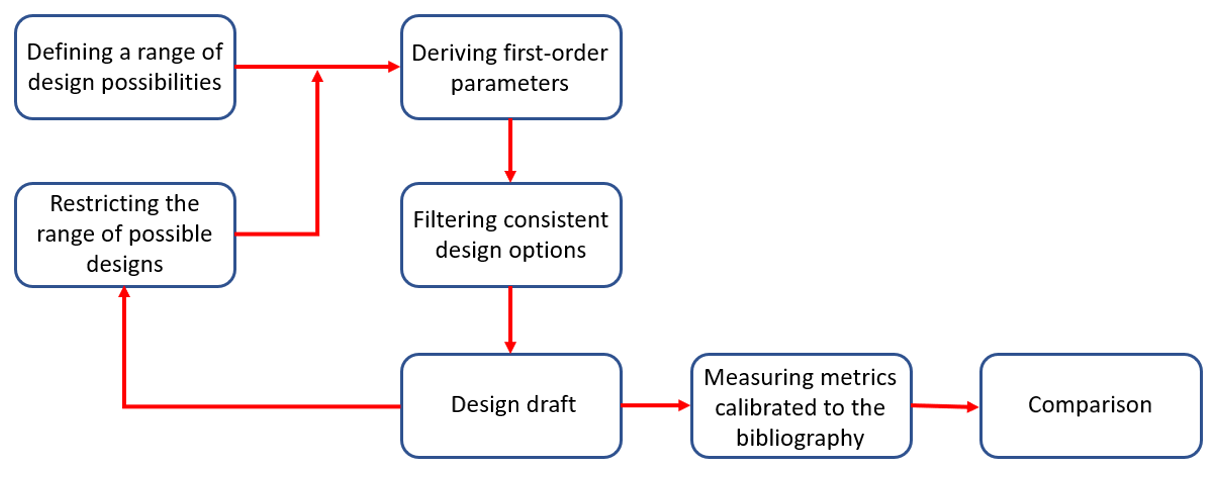}
\end{tabular}
\end{center}
\caption[SC] 
{ \label{fig:flowchart} 
IFS spectrographs trade-off flowchart.}
\end{figure}

The flowchart shown in Fig.~\ref{fig:flowchart} summarises the preparatory work carried out prior to selecting an architecture. A key question running throughout this analysis is whether curved detectors are worth adopting, particularly in terms of throughput, cost, and risk; this is addressed in detail in Sec.~\ref{sec:designs}.

\subsection{Geometrical etendue and number of spectrographs}
\label{sec:etendue}

The characteristics of the design options explored in this study fall into three categories: spectrograph camera design, detector characteristics and sampling. Two camera designs have been identified: dioptric and catadioptric. Eight detector options are considered, as listed in Tab.~\ref{tab:list}.

\begin{table}[H]
\caption{Characteristics of the detectors considered in this study.} 
\label{tab:list}
\begin{center}       
\begin{tabular}{|>{\centering\arraybackslash}m{1.5cm}|>{\centering\arraybackslash}m{2.4cm}|>{\centering\arraybackslash}m{1.8cm}|>{\centering\arraybackslash}m{2.2cm}|>{\centering\arraybackslash}m{3.5cm}|} 
\hline
\rule[-1ex]{0pt}{3.5ex} Name & \shortstack{Format\\$[\mathrm{pixels}\times\mathrm{pixels}]$} & \shortstack{pixel size\\$[\mathrm{\mu m}]$} & \shortstack{Detector size\\$[\mathrm{mm}\times \mathrm{mm}]$} & Binning  \\
\hline
\rule[-1ex]{0pt}{3.5ex}  4k-15µm & 4000$\times$4000 & 15 & 60$\times$60 & None   \\
\hline
\rule[-1ex]{0pt}{3.5ex}  6k-15µm & 6000$\times$6000 & 15 & 90$\times$90 & None   \\
\hline
\rule[-1ex]{0pt}{3.5ex}  6k-10µm & 6000$\times$6000 & 10 & 60$\times$60 & None   \\
\hline
\rule[-1ex]{0pt}{3.5ex}  6k-10µm & 6000$\times$6000 & 10 & 60$\times$60 & x2 along spatial axis   \\
\hline
\rule[-1ex]{0pt}{3.5ex}  8k-10µm & 8000$\times$8000 & 10 & 80$\times$80 & x2 along spatial axis   \\
\hline
\rule[-1ex]{0pt}{3.5ex}  9k-10µm & 9000$\times$9000 & 10 & 90$\times$90 & x2 along spatial axis   \\
\hline
\rule[-1ex]{0pt}{3.5ex}  9k-10µm & 9000$\times$9000 & 10 & 90$\times$90 & x3 along spatial axis   \\
\hline
\rule[-1ex]{0pt}{3.5ex}  6k-15µm & 6000$\times$6000 & 15 & 90$\times$90 & x2 along spatial axis   \\
\hline
\end{tabular}
\end{center}
\end{table}

All detectors listed are square, though custom rectangular formats could be explored as a means to lower the total pixel count in the instrument. For the spatial sampling, two options are considered: a square 0.25"$\times$0.25" sampling and a rectangular 0.3"$\times$0.2" sampling. The latter is less well suited to the instrument since better seeing statistics are expected with Ground Layer Adaptive Optics, making a 0.3" spaxel an inefficient match; nevertheless, it illustrates the impact that non-square sampling can have on the instrument's design.

\begin{figure} [H]
\begin{center}
\begin{tabular}{c} 
\includegraphics[height=7cm]{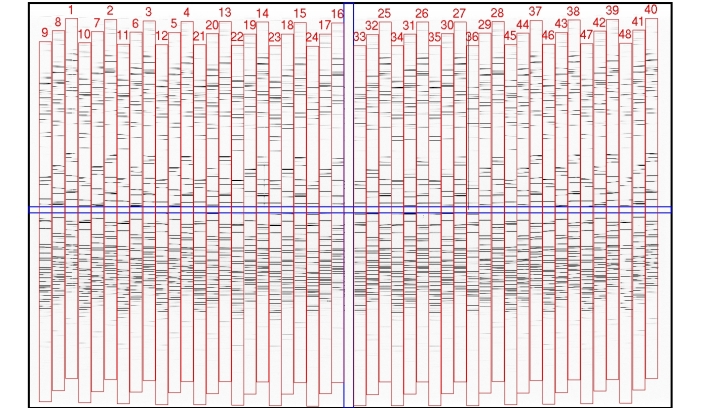}
\end{tabular}
\end{center}
\caption[SC] 
{ \label{fig:MUSE_det} 
Arrangement of the spectra of the 48 slitlets on the MUSE detectors.}
\end{figure}

The entrance slit of a spectrograph fed by an image slicer has an unusual geometry: it consists of a series of slightly spaced slitlets arranged in a stepped pattern (see Fig.~\ref{fig:MUSE_det}). As a result, neither the spatial nor the spectral coverage spans the full detector area. Some spaxels fall in the gaps between slitlets, while the common spectral band covers slightly less than the full detector width along the dispersion axis. Throughout this document, we therefore assume that 90\% of the detector pixels and spaxels are effectively used by the spectrograph, consistent with what was achieved on MUSE and estimated for BlueMUSE.

The choice of detector and sampling is fundamental, as it determines the number of spectrographs in the instrument, the grating characteristics, and the aperture and field of view of each individual unit, and therefore the overall volume and cost of the instrument. The remainder of this section quantifies the impact of these choices on critical design parameters.

\begin{figure} [H]
\begin{center}
\begin{tabular}{c} 
\includegraphics[height=5.5cm]{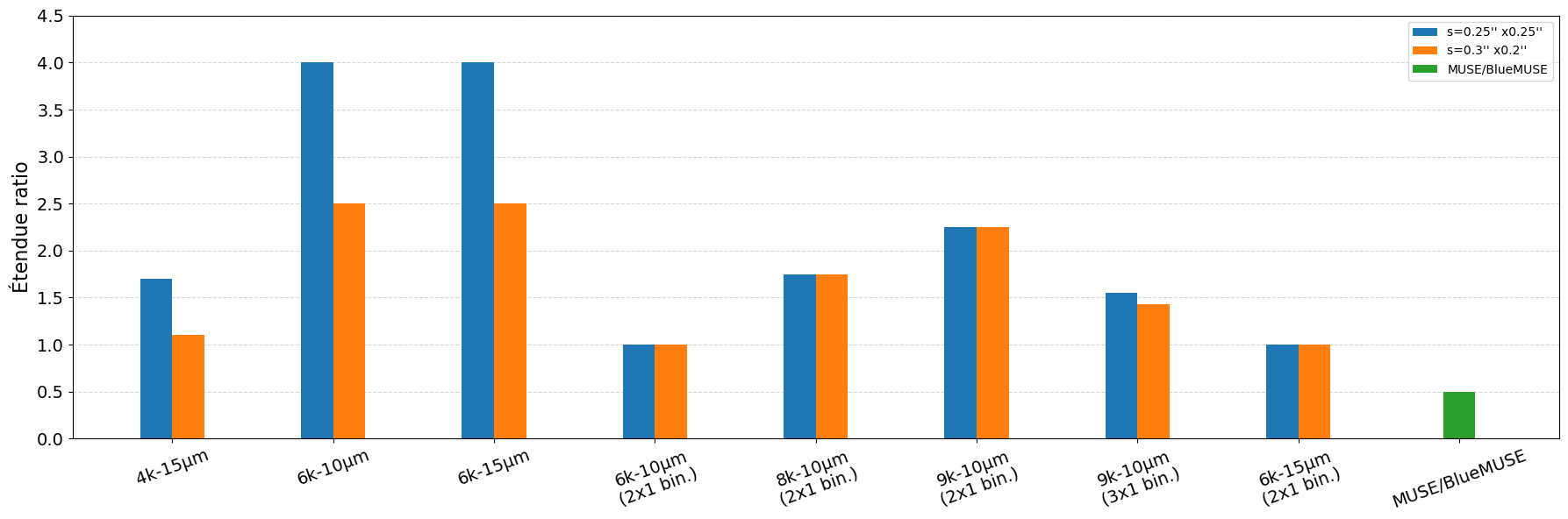}
\end{tabular}
\end{center}
\caption[SC] 
{ \label{fig:GeomEtendue} 
Geometric étendue of the camera for the different detector options, normalised to the minimum value obtained for the 6k-10µm detector with square sampling.}
\end{figure}

\newpage

Fig.~\ref{fig:GeomEtendue} illustrates the geometric etendue ($\pi\left(\frac{D_{det}}{2f_/}\right)^2$) required by each design option, normalized to the 6k-10µm detector with binning and square sampling. While etendue does not fully capture optical design complexity, it provides a useful first-order indicator: larger etendue generally implies a more demanding camera design, requiring more optical surfaces and stronger aspherizations, which in turn reduces throughput, increases volume, and raises cost. As an example, for equal spatial sampling, the 6k-15µm detector yields a more complex camera than the binned 9k-10µm option, which is itself more demanding than the 4k-15µm. Tab.~\ref{tab:n_spectro} further shows that a rectangular sampling does not significantly increase the number of spectrographs required --- true as long as the geometric mean of the sampling stays similar to that of the square sampling --- though for some configurations it can reduce the camera étendue.

\begin{table}[H]
\caption{Number of spectrographs and camera aperture for the 
detector options considered.} 
\label{tab:n_spectro}
\begin{center}       
\begin{tabular}{|>{\centering\arraybackslash}m{2.2cm}|>{\centering\arraybackslash}m{2.4cm}|>{\centering\arraybackslash}m{2.4cm}|>{\centering\arraybackslash}m{2.4cm}|>{\centering\arraybackslash}m{2.4cm}|} 
\hline
\rule[-1ex]{0pt}{3.5ex}  & \multicolumn{2}{c|}{Number of spectrographs} & \multicolumn{2}{c|}{Camera aperture f/}   \\
\cline{2-5}
\rule[-1ex]{0pt}{3.5ex}   & 0.25"x0.25" & 0.3"x0.2" & 0.25"x0.25" & 0.3"x0.2"   \\
\hline
\rule[-1ex]{0pt}{3.5ex}  4k-15µm & 144 & 150 & 1.03 & 1.29  \\
\hline
\rule[-1ex]{0pt}{3.5ex}  6k-10µm & 96 & 100 & 0.69 & 0.86  \\
\hline
\rule[-1ex]{0pt}{3.5ex}  6k-15µm & 96 & 100 & 1.03 & 1.29  \\
\hline
\rule[-1ex]{0pt}{3.5ex}  \shortstack{6k-10µm\\(2x1 bin.)} & 192 & 200 & 1.37 & 1.37  \\
\hline
\rule[-1ex]{0pt}{3.5ex}  \shortstack{8k-10µm\\(2x1 bin.)} & 144 & 150 & 1.37 & 1.37  \\
\hline
\rule[-1ex]{0pt}{3.5ex}  \shortstack{9k-10µm\\(2x1 bin.)} & 128 & 134 & 1.37 & 1.37  \\
\hline
\rule[-1ex]{0pt}{3.5ex}  \shortstack{9k-10µm\\(3x1 bin.)} & 192 & 200 & 1.65 & 1.72  \\
\hline
\rule[-1ex]{0pt}{3.5ex}  \shortstack{6k-15µm\\(2x1 bin.)} & 192 & 200 & 2.06 & 2.06  \\
\hline

\end{tabular}
\end{center}
\end{table}

Together, Fig.~\ref{fig:GeomEtendue} and Tab.~\ref{tab:n_spectro} highlight a central tension in the design space: architectures with the smallest etendue per spectrograph have the simplest camera designs but require the largest number of units. Conversely, increasing the etendue per spectrograph reduces the unit count but raises individual complexity. Whether it is more cost-effective to build many simple spectrographs and exploit the serial effect, or fewer complex ones, is one of the central questions of this trade-off and is addressed quantitatively in Sec.~\ref{sec: cost}.

\subsection{Anamorphism ratio}

Anamorphism is an important design degree of freedom in integral field spectrographs: compressing the beam differently along the spatial and spectral axes ensures correct sampling by the spectrograph. Hereafter we define the anamorphism ratio as the ratio of the pupil diameter along the spatial axis to that along the spectral axis:

\begin{equation}
A = \frac{s}{b}\frac{\theta_{s,x}}{\theta_{s,y}},
\label{eq:Anamorphism}
\end{equation}

\noindent where $A$ is the anamorphism ratio, $s$ is the slit width in pixels, $b$ is the binning factor (ratio of spaxel size to pixel size), and $\theta_{s,x}$, $\theta_{s,y}$ are the spatial samplings along the spatial and spectral axes respectively. Note that $s$ is not the line spread function; the LSF is the convolution of $s$ with the combined image quality contribution from the image slicer entrance to the detector. Also, this expression assumes the spectrograph operates in Littrow configuration. 

Applying Eq.~\ref{eq:Anamorphism} recovers $A = 2$ for MUSE and $A = 1.33$ for BlueMUSE, consistent with the known instrument parameters. The values obtained for the WST IFS design options are shown in Tab.~\ref{tab:A_list}; the slit sizes either correspond to the maximum permissible values to remain compliant with the current spectral resolution requirements, or to a deliberately capped value of 2.4 pixels.

\begin{table}[H]
\caption{Anamorphism ratio for the WST IFS design options.} 
\label{tab:A_list}
\begin{center}       
\begin{tabular}{|>{\centering\arraybackslash}p{3cm}|>{\centering\arraybackslash}p{1.8cm}|>{\centering\arraybackslash}p{1.8cm}|>{\centering\arraybackslash}p{1.8cm}|>{\centering\arraybackslash}p{1.8cm}|>{\centering\arraybackslash}p{1.8cm}|} 
\hline
\rule[-1ex]{0pt}{3.5ex}  Detector & s [pixels] & b & $\theta_{s,x}$ [arcsec] & $\theta_{s,y}$ [arcsec]& A  \\
\hline
\rule[-1ex]{0pt}{3.5ex}  \multirow{2}{*}{4k-15µm} & \multirow{2}{*}{2} & \multirow{2}{*}{1} & 0.25 & 0.25 & 2  \\
\cline{4-6}
\rule[-1ex]{0pt}{3.5ex}   &  &  & 0.2 & 0.3 & 1.33   \\
\hline
\rule[-1ex]{0pt}{3.5ex}  \multirow{2}{*}{6k-10µm} & \multirow{2}{*}{2.4} & \multirow{2}{*}{1} & 0.25 & 0.25 & 2.4   \\
\cline{4-6}
\rule[-1ex]{0pt}{3.5ex}   &  &  & 0.2 & 0.3 & 1.6   \\
\hline
\rule[-1ex]{0pt}{3.5ex}  \multirow{2}{*}{6k-15µm} & \multirow{2}{*}{2.4} & \multirow{2}{*}{1} & 0.25 & 0.25 & 2.4   \\
\cline{4-6}
\rule[-1ex]{0pt}{3.5ex}   &  &  & 0.2 & 0.3 & 1.6   \\
\hline
\rule[-1ex]{0pt}{3.5ex}  \multirow{2}{*}{\shortstack{6k-10µm\\(2x1 bin.)}} & \multirow{2}{*}{2.4} & \multirow{2}{*}{2} & 0.25 & 0.25 & 1.2   \\
\cline{4-6}
\rule[-1ex]{0pt}{3.5ex}   &  &  & 0.2 & 0.3 & 0.8   \\
\hline
\rule[-1ex]{0pt}{3.5ex}  \multirow{2}{*}{\shortstack{8k-10µm\\(2x1 bin.)}} & \multirow{2}{*}{2.4} & \multirow{2}{*}{2} & 0.25 & 0.25 & 1.2   \\
\cline{4-6}
\rule[-1ex]{0pt}{3.5ex}   &  &  & 0.2 & 0.3 & 0.8   \\
\hline
\rule[-1ex]{0pt}{3.5ex}  \multirow{2}{*}{\shortstack{9k-10µm\\(2x1 bin.)}} & \multirow{2}{*}{2.4} & \multirow{2}{*}{2} & 0.25 & 0.25 & 1.2   \\
\cline{4-6}
\rule[-1ex]{0pt}{3.5ex}   &  &  & 0.2 & 0.3 & 0.8   \\
\hline
\rule[-1ex]{0pt}{3.5ex}  \multirow{2}{*}{\shortstack{9k-10µm\\(3x1 bin.)}} & \multirow{2}{*}{2.4} & \multirow{2}{*}{3} & 0.25 & 0.25 & 0.8   \\
\cline{4-6}
\rule[-1ex]{0pt}{3.5ex}   &  &  & 0.3 & 0.2 & 1.2   \\
\hline
\rule[-1ex]{0pt}{3.5ex}  \multirow{2}{*}{\shortstack{6k-15µm\\(2x1 bin.)}} & \multirow{2}{*}{2.4} & \multirow{2}{*}{2} & 0.25 & 0.25 & 1.2   \\
\cline{4-6}
\rule[-1ex]{0pt}{3.5ex}   &  &  & 0.2 & 0.3 & 0.8   \\
\hline
\end{tabular}
\end{center}
\end{table}

When $A > 1$, the pupil is wider along the spatial axis than the spectral axis, which is the more favorable configuration: the spectral axis presents a slower focal ratio, slightly relaxing the camera design constraints. When $A < 1$, the situation is reversed: the camera must be optimized for a faster focal ratio than what the spatial sampling alone would require, which increases design complexity. This regime arises for several binned configurations and can represent a non-trivial constraint on camera optimization. It could be prevented for some rectangular sampling options if the values of $\theta_{s,y}$ and $\theta_{s,x}$ are reversed, but this would lead to even smaller f-numbers.

\subsection{Pupil diameter estimation}
\label{sec:pupil}

The pupil diameter is a key design parameter, directly influencing transmission, the camera's half-field of view, the physical volume of the spectrograph, and its cost. Estimating this diameter ahead of any full optical design allows us to bound the parameter space and assign physically motivated starting values to draft designs. Because the instrument's anamorphism produces an elliptical pupil, the term "pupil diameter" throughout this document refers to the larger of the two axes of the ellipse. Accordingly, the focal ratio always refers to the faster of the two axes.

Rather than deriving the pupil diameter from first principles, we adopt an empirical power law calibrated on a representative sample of existing spectrograph cameras; the full calibration procedure, instrument list, and fit diagnostics are given in Appendix~\ref{app:pupil}. The underlying assumption is that WST spectrographs will be designed with broadly similar engineering choices to past instruments: comparable optical complexity, similar glass catalogues and constraints, and similar image quality targets. The approach is not intended to predict the exact pupil diameter of a specific optimised design, but to give a representative order-of-magnitude estimate that can guide early trade-off decisions. For dioptric designs, the fitted relation is:

\begin{equation}
D_{pupil} = \alpha(A\Omega^{\delta})^{\beta} = \alpha(D_{det}^2(\frac{\pi}{4{f_/}^2})^{\delta})^{\beta},
\label{eq:Dpup_mod}
\end{equation}

\noindent with $\alpha = 9.7163$~a.u., $\beta = 0.4434$, $\delta = 1.75$ ($R^2 = 0.958$). For catadioptric designs, the central obstruction imposed by the detector is amplified by the fore-optics anamorphism, which compresses the pupil along one axis. The calibration variable is therefore augmented by the anamorphism ratio $\gamma$:

\begin{equation}
D_{pupil} = \alpha(\gamma A\Omega^{\delta})^{\beta} = \alpha(\gamma D_{det}^2(\frac{\pi}{4{f_/}^2})^{\delta})^{\beta},
\label{eq:Dpup_cata}
\end{equation}

\noindent with $\alpha = 18.329$~a.u., $\beta = 0.3436$, $\delta = 1.3$ ($R^2 = 0.922$). The exponent $\delta$ carries no direct physical interpretation; it simply reflects the difference in correlation structure with aperture and field present in the calibration sample. The resulting pupil diameter estimates for all design options are given in Tab.~\ref{tab:Dpup}.

\begin{table}[H]
\caption{Estimated pupil diameter for all dioptric (left) and catadioptric (right) design options and the two samplings.}
\label{tab:Dpup}
\begin{center}
\begin{tabular}{|>{\centering\arraybackslash}m{3.5cm}|
                 >{\centering\arraybackslash}m{2.5cm}|
                 >{\centering\arraybackslash}m{2.5cm}|
                 >{\centering\arraybackslash}m{2.5cm}|
                 >{\centering\arraybackslash}m{2.5cm}|}
\hline
\rule[-1ex]{0pt}{3.5ex} Detector 
    & \multicolumn{2}{c|}{Dioptric pupil diameter [mm]} 
    & \multicolumn{2}{c|}{Catadioptric pupil diameter [mm]} \\
\cline{2-5}
\rule[-1ex]{0pt}{3.5ex} & 0.25"$\times$0.25" 
                        & 0.3"$\times$0.2"
                        & 0.25"$\times$0.25"
                        & 0.3"$\times$0.2" \\
\hline
\rule[-1ex]{0pt}{3.5ex} 4k-15µm              & 290 & 205 & 325 & 230 \\
\hline
\rule[-1ex]{0pt}{3.5ex} 6k-10µm              & 540 & 385 & 495 & 355 \\
\hline
\rule[-1ex]{0pt}{3.5ex} 6k-15µm              & 415 & 290 & 460 & 325 \\
\hline
\rule[-1ex]{0pt}{3.5ex} 6k-10µm (2$\times$1 bin.) & 185 & 185 & 210 & 185 \\
\hline
\rule[-1ex]{0pt}{3.5ex} 8k-10µm (2$\times$1 bin.) & 240 & 240 & 260 & 225 \\
\hline
\rule[-1ex]{0pt}{3.5ex} 9k-10µm (2$\times$1 bin.) & 270 & 270 & 280 & 240 \\
\hline
\rule[-1ex]{0pt}{3.5ex} 9k-10µm (3$\times$1 bin.) & 200 & 190 & 205 & 230 \\
\hline
\rule[-1ex]{0pt}{3.5ex} 6k-15µm (2$\times$1 bin.) & 140 & 140 & 195 & 170 \\
\hline
\end{tabular}
\end{center}
\end{table}

Estimates at the upper end of the étendue range lie well outside the calibration domain and should be treated with caution, as extrapolation uncertainty grows accordingly. It is also worth noting that for the catadioptric designs, the largest estimated diameters (6k-10µm and 6k-15µm without binning) assume a slit sampling of 2.4~pixels; reducing this to 2~pixels would decrease the anamorphism ratio and, provided $A > 1$, reduce the estimated pupil diameter.

\subsection{Spectral resolution and sampling}

\begin{figure} [H]
\begin{center}
\begin{tabular}{c} 
\includegraphics[height=7.0cm]{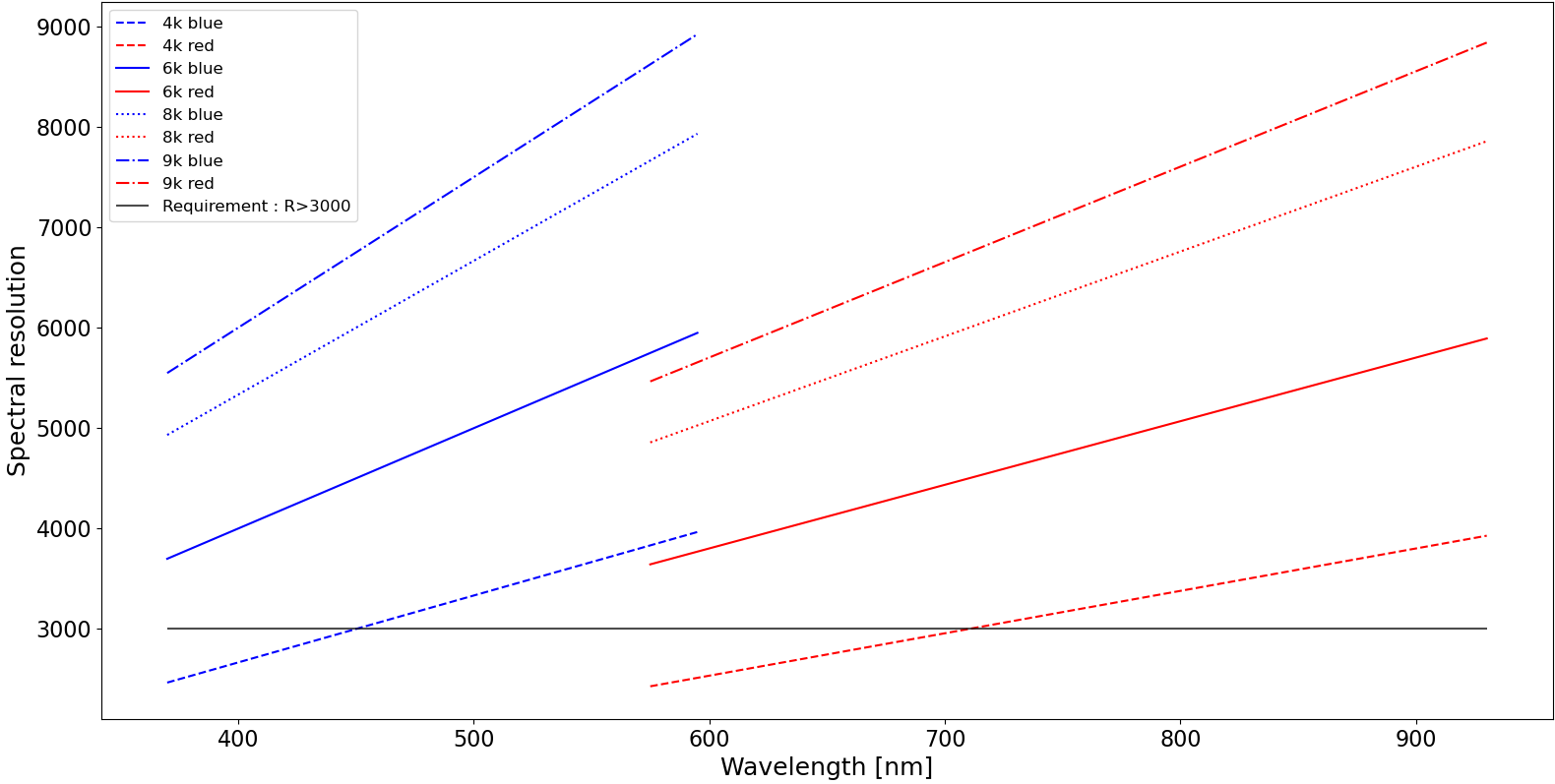}
\end{tabular}
\end{center}
\caption[SC] 
{ \label{fig:Res_wave} 
Spectral resolution across both arms for the four pixel counts along the dispersion axis, assuming a spectral sampling of 2.4.}
\end{figure}

Another key parameter largely influenced by the detector characteristics is the spectral resolution. Fig.~\ref{fig:Res_wave} shows the spectral resolution across both arms for four values of the pixel count, assuming a slit sampling of 2.4 pixels. The blue arm covers 370--595~nm and the red arm 575--930~nm; these boundaries can be fine-tuned at a later stage, but already give a clear picture of the trade-offs involved, as well as the degree of flexibility available.

\begin{figure} [H]
\begin{center}
\begin{tabular}{c} 
\includegraphics[height=7cm]{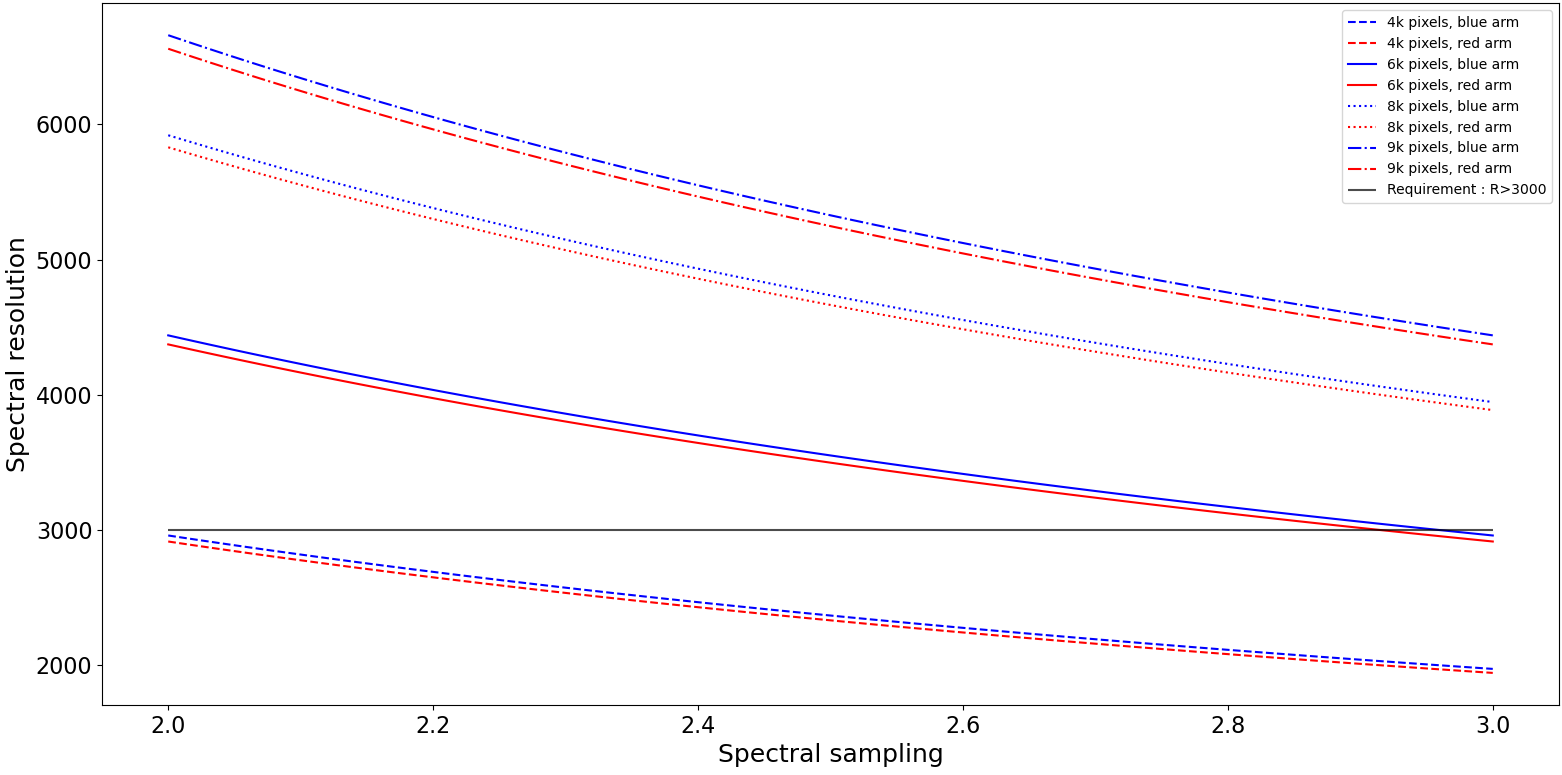}
\end{tabular}
\end{center}
\caption[SC] 
{ \label{fig:Res_samp} 
Spectral resolution at 370~nm (blue channel) and 575~nm (red channel) as a function of the spectral sampling of the slit.}
\end{figure}

The 4k detector falls well short of the R$>$3000 requirement at a sampling of 2.4 pixels, and can only be made compliant by reducing the sampling to around 2.0--2.1 pixels (see Fig.~\ref{fig:Res_samp}), which leaves little margin for adjustment. At the other extreme, the 8k and 9k detectors exceed the requirements comfortably, but at the cost of a much larger pixel count. A 6k detector represents a practical middle ground: it satisfies the resolution requirement with enough margin to allow the spectral sampling to be increased if needed, while keeping the total pixel count manageable. Indeed, as Fig.~\ref{fig:Res_samp} illustrates, even 5k pixels along the spectral axis would be sufficient to meet the resolution requirement at a sampling of 2.4 pixels.

%% file: 3-Evaluation_metrics.tex
\newpage
\section{EVALUATION METRICS}
\label{sec:metrics}

This section presents the metrics used to compare the design options and, where relevant, to filter out configurations that are not viable. We distinguish between performance metrics, which bear directly on the scientific output of the instrument, and engineering constraints, which set practical boundaries on what can be built and operated. Together, these metrics feed into the design space down-selection carried out in Sec.~\ref{sec:designs}.

\subsection{Performance metrics}
\label{sec:Perf}

\subsubsection*{Error budget breakdown for the spectrographs and detectors}

A fundamental constraint on the camera design is the image quality budget: the allowed contribution of the spectrograph and detector to the seeing degradation, when projected onto the sky. We assess this budget separately for the detector and the spectrograph optics, as the two contributions scale differently with camera speed.

\begin{figure} [H]
\begin{center}
\begin{tabular}{c} 
\includegraphics[height=6.5cm]{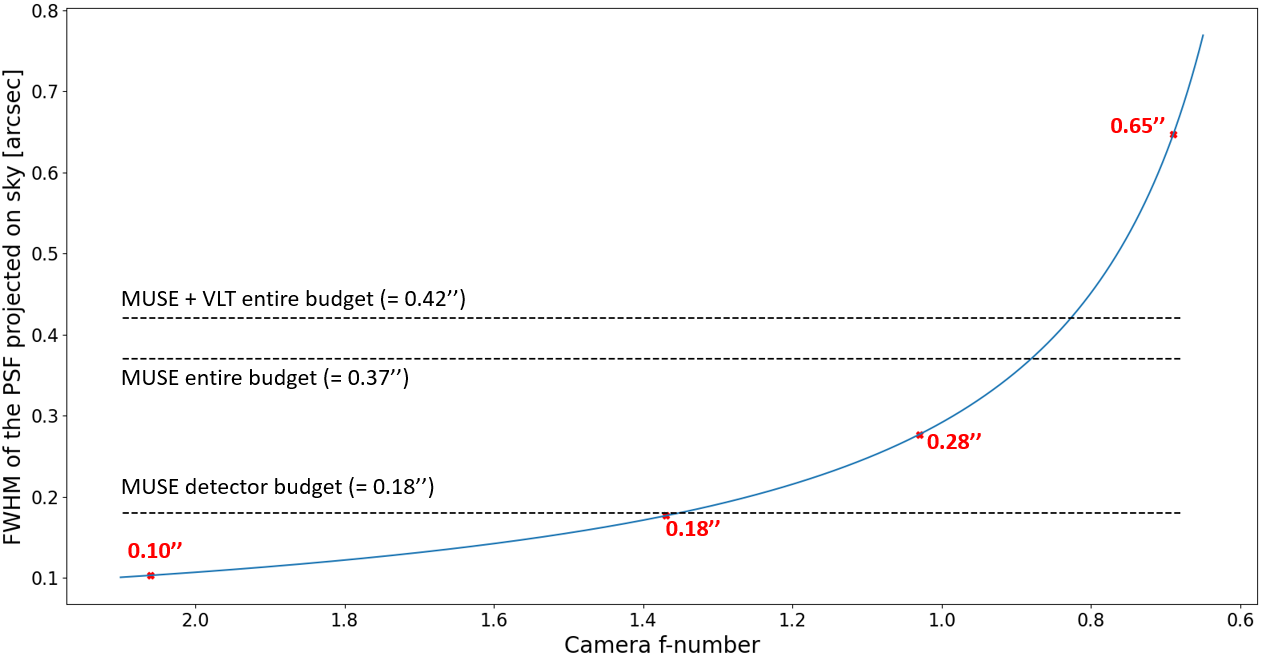}
\end{tabular}
\end{center}
\caption[SC] 
{ \label{fig:det_budget} 
Detector error budget (PSF, flatness and positioning) projected on sky, as a function of the camera f-number.}
\end{figure}

For the detector, three terms contribute to image degradation: surface flatness, positioning errors, and the intrinsic detector PSF. We adopt the error budgets established for the BlueMUSE detectors, summarised in Tab.~\ref{tab:det_budget}, as a reference. These three terms are combined in quadrature and projected onto the sky in Fig.\ref{fig:det_budget}, without including seeing or any other instrument subsystem.

\begin{table}[H]
\caption{Detector error budget taken from BlueMUSE.} 
\label{tab:det_budget}
\begin{center}       
\begin{tabular}{|>{\centering\arraybackslash}p{4.5cm}|>{\centering\arraybackslash}p{2.5cm}|>{\centering\arraybackslash}p{2.5cm}|>{\centering\arraybackslash}p{2.5cm}|} 
\hline
\rule[-1ex]{0pt}{3.5ex}   & Flatness & Positioning & PSF  \\
\hline
\rule[-1ex]{0pt}{3.5ex}  Defocus [µm] & $\pm$10 PV & $\pm$5 & N/A   \\
\hline
\rule[-1ex]{0pt}{3.5ex}  FWHM on the detector [µm] & Depends on f/ & Depends on f/ & 11    \\
\hline
\end{tabular}
\end{center}
\end{table}

The detector budget grows rapidly as the camera becomes faster. A camera at f/1.35 already allocates a budget to the detector comparable to that of MUSE, which is known to be demanding. Below f/1.35, the budget increases sharply: at f/1.03, maintaining the MUSE flatness specification of 10~$\mu$m half-PV would push the whole detector contribution to 0.28" FWHM on sky. At f/0.7, the detector alone would consume more than the entire MUSE instrument budget. Conversely, if the camera is slowed down to a value greater than f/1.35, the budget allocated to the detector will be lower when projected onto the sky. At f/2.06, this budget is only 0.10" FWHM, compared to 0.18" FWHM for MUSE. This places a clear lower bound on the acceptable camera f-number: designs faster than f/1.35 will require either relaxing the detector flatness specification, accepting a larger detector error budget and image degradation, or tightening the budgets allocated elsewhere.

\begin{figure} [H]
\begin{center}
\begin{tabular}{c} 
\includegraphics[height=6.5cm]{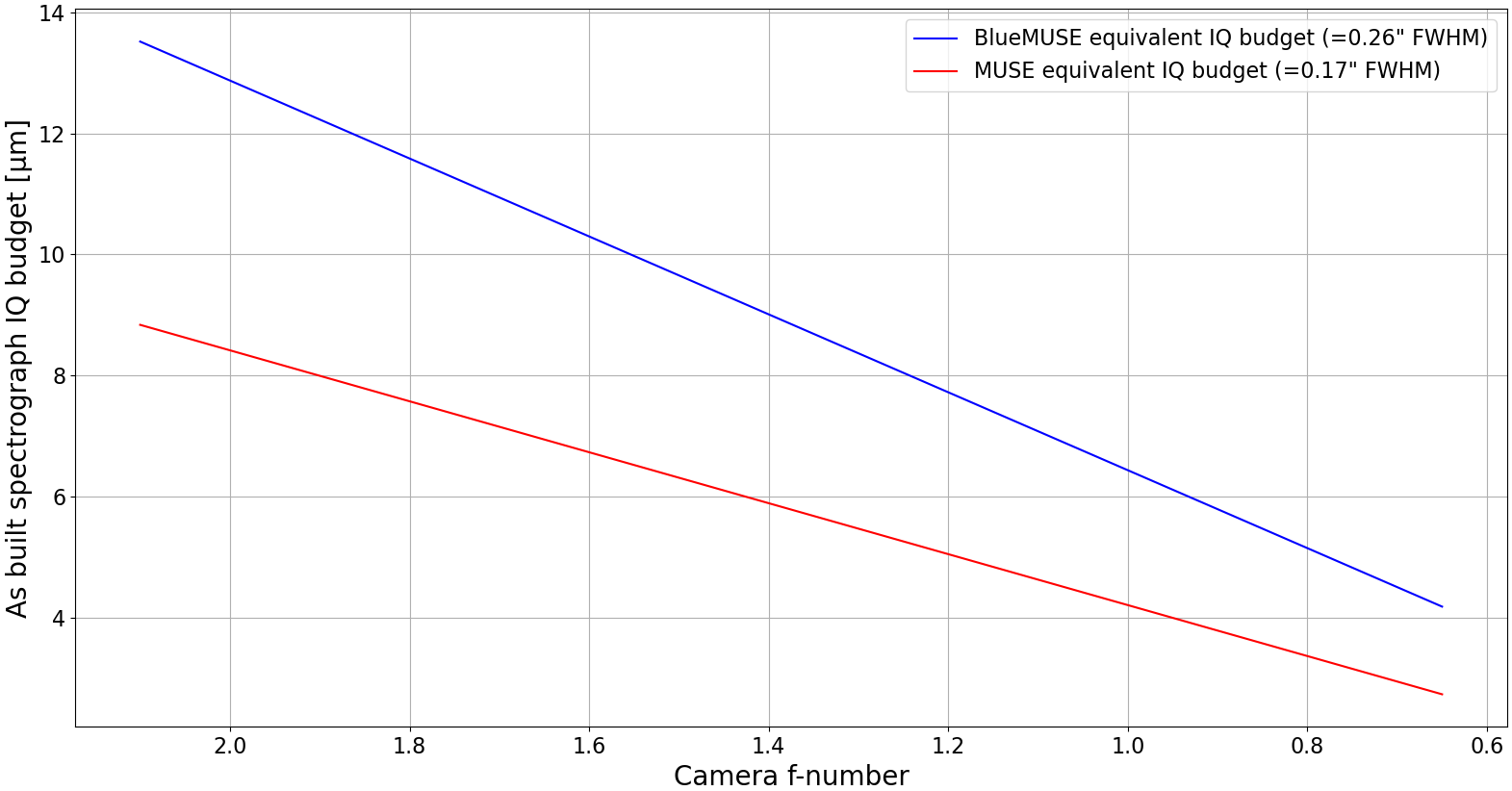}
\end{tabular}
\end{center}
\caption[SC] 
{ \label{fig:IQ_budget} 
As-built MUSE and BlueMUSE spectrograph image quality budget as a function of camera f-number.}
\end{figure}

The spectrograph optics budget is equally affected by camera speed, through two compounding effects. First, a faster camera is intrinsically harder to design: correcting aberrations over a wide field at small f-number generally requires many surfaces and strong aspherisations. Second, as Fig.~\ref{fig:IQ_budget} shows, the physical image quality budget allocated to the spectrograph decreases as the camera gets faster, since the same angular budget on sky corresponds to a smaller spot on the detector. Consequently, a fast camera is not only more complex to design, but must also achieve intrinsically better performance, with tighter tolerances throughout.

Taken together, these two considerations strongly argue against cameras faster than approximately f/1.35 for WST. This threshold is used in Sec.~\ref{sec:designs} to filter out design options whose camera speed makes them impractical.

\subsubsection*{Throughput: dioptric transmission against catadioptric obstruction}

Given the extensive number of optical surfaces expected, one of the primary design drivers is to maximise the throughput of the spectrographs across the full wavelength range. Whether a dioptric or catadioptric architecture is preferable from a throughput standpoint depends on a trade-off between the transmission losses introduced by the lenses of a dioptric design, and the light loss caused by the central obstruction of a catadioptric one. The relevant question can therefore be stated simply: for a given level of additional obstruction in a catadioptric camera, how many lenses can a dioptric design accommodate before its transmission drops below that of the catadioptric alternative?

We note that "additional obstruction" here refers only to the obstruction introduced within the spectrograph itself, over and above that already present from the telescope baffle.

Starting from a simplified model of the total camera transmission,

\begin{equation}
T(x, \lambda) = T_{obs}(x,\lambda)\prod_{lenses}T_{in}(\lambda)T_{out}(\lambda)e^{-\alpha(\lambda)d}\prod_{mirrors}R(\lambda),
\label{eq:T_full}
\end{equation}

\noindent where $x$ is the field position on the slit, $T_{obs}$ is the fraction of light unobstructed in the spectrograph, $T_{in}$ and $T_{out}$ are the coating efficiencies on the two sides of a lens, $e^{-\alpha(\lambda)d}$ corresponds to the fraction of light unabsorbed by a lens of thickness $d$, and $R$ is the coating efficiency of a mirror.

Assuming that obstruction is wavelength- and field-independent, that each coating has a common transmission $T_\mathrm{in}(\lambda) = T_\mathrm{out}(\lambda) = R(\lambda) = 0.993$, and the transmission of a set of $n$ lenses can be approximated by an equivalent per-lens transmission $T_\mathrm{eq}$ raised to the power $n$, yielding a simplified expression:

\begin{equation}
T = T_{obs}T_{eq}^nR^m,
\label{eq:T_simple}
\end{equation}

\noindent where $n$ is the number of lenses and $m$ the number of mirrors. Three values of $T_\mathrm{eq}$ are retained to bracket the uncertainty in glass choice. They correspond to the highest, average, and lowest transmission estimated over a representative set of simulated lenses in the materials SFSL5Y, BAL35Y, PBL35Y, PBL1Y, Silica, and CaF$_2$ with thicknesses ranging from 20 to 80 mm. Assuming $n_0 = 3$ lenses in the catadioptric design (consistent with the 2--4 lenses typical of such cameras) and setting the dioptric and catadioptric transmissions equal gives the maximum number of lenses permissible in an equivalent dioptric design:

\begin{equation}
n = n_0 + \frac{log(T_{obs}R)}{log(T_{eq})},
\label{eq:n_lenses}
\end{equation}

\begin{figure} [H]
\begin{center}
\begin{tabular}{c} 
\includegraphics[height=6.5cm]{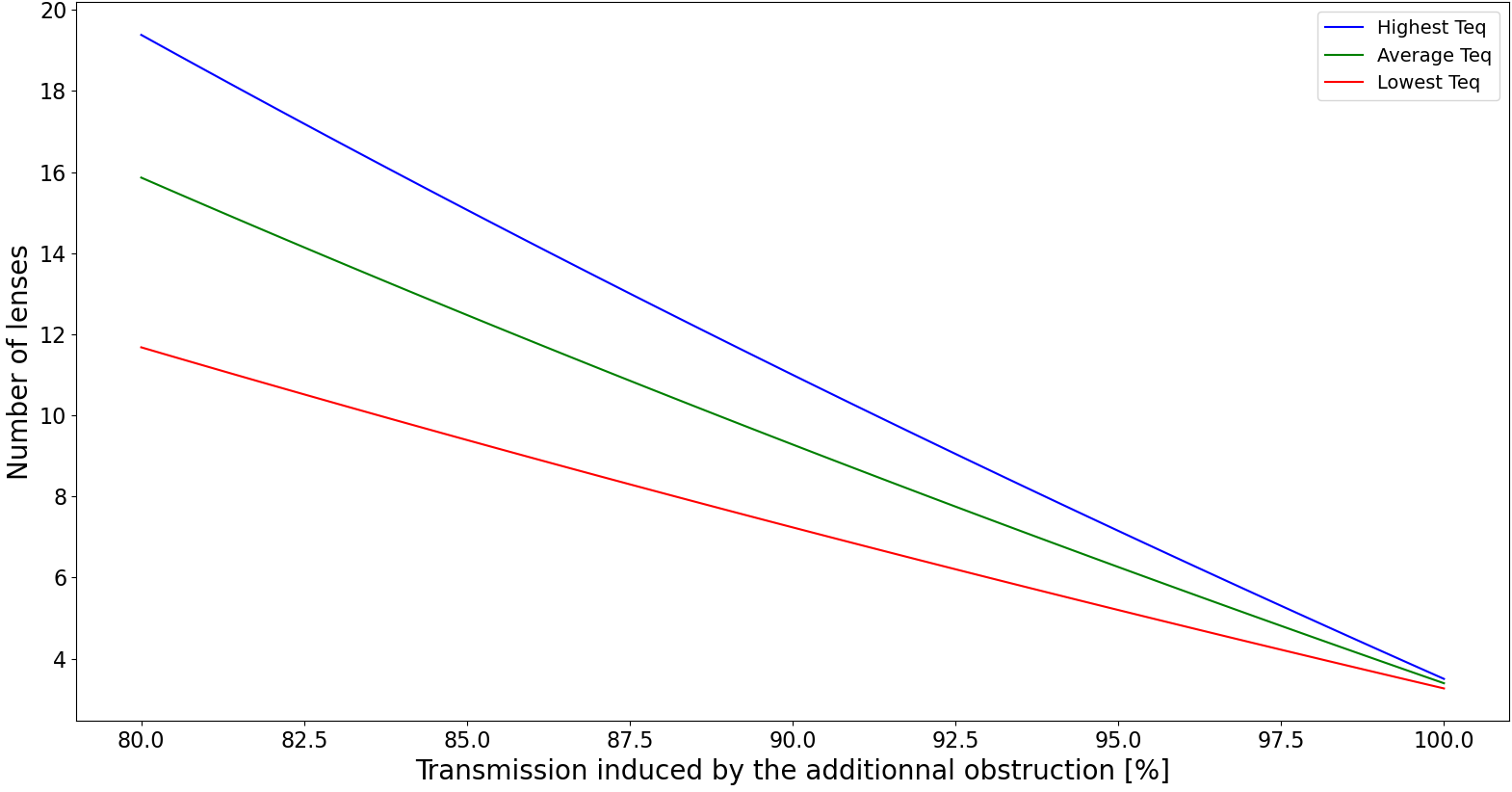}
\end{tabular}
\end{center}
\caption[SC] 
{ \label{fig:nlens_obs} 
Maximum number of lenses in a dioptric design yielding the same throughput as a catadioptric design with a given level of additional obstruction.}
\end{figure} 

The result is shown in Fig.~\ref{fig:nlens_obs} as a function of the transmission induced by the additional obstruction ($T_\mathrm{obs}$). In the average case, a 10\% additional obstruction ($T_\mathrm{obs} = 90\%$) allows up to 9 lenses in a dioptric design; with low-absorption materials such as Silica or CaF$_2$, this rises to approximately 11.

In practice, the additional obstruction is more naturally expressed in terms of the pupil diameter rather than as a fractional area, since the former is the quantity directly available from the scaling relations of Sec.~\ref{sec:pupil}. The dependence of obstruction on pupil diameter, detector margin, and anamorphism ratio is derived in Appendix~\ref{app:obstruction}. The key result is that a small change in the margin around the detector has a large effect on the additional obstruction at small pupil diameters, and that the anamorphism ratio plays a decisive role: the compression of the pupil by a factor $\gamma$ can increase the additional obstruction significantly if the pupil diameter is not large enough. As an illustration, with a 5 mm margin around a 60 mm sensor and $\gamma = 2$, the pupil must exceed approximately 260 mm to keep the extra obstruction at or below 10\%.

\begin{figure} [H]
\begin{center}
\begin{tabular}{c} 
\includegraphics[height=12.0cm]{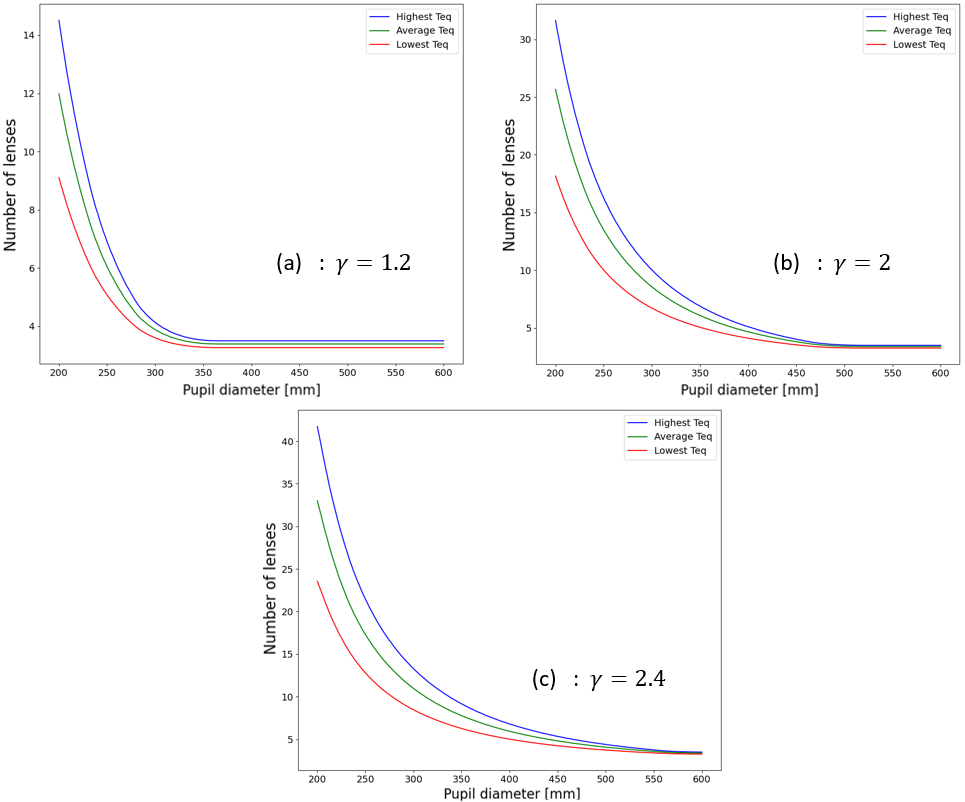}
\end{tabular}
\end{center}
\caption[SC] 
{ \label{fig:nlens_pup} 
Maximum number of lenses in a dioptric design before a catadioptric design becomes preferable on throughput grounds, as a function of pupil diameter. The additional obstruction is that of a disc enclosing a 60~mm detector with a 10~mm margin. Each panel corresponds to a fixed anamorphism ratio: $\gamma = 1.2$ (a), $\gamma = 2$ (b), and $\gamma = 2.4$ (c).}
\end{figure}

Combining this result with Fig.~\ref{fig:nlens_obs} gives Fig.~\ref{fig:nlens_pup}, which directly expresses the maximum permissible number of lenses in a dioptric design as a function of pupil diameter, for a 60 mm detector with a 10 mm margin and three anamorphism ratios. This figure is the key throughput filter for the design trade-off: for detectors larger than 60 mm, the additional obstruction of a catadioptric design becomes prohibitive at any realistic pupil size, and a dioptric design is strongly preferred. For a 60 mm detector, the conclusion depends on the pupil size and anamorphism ratio. For instance, with $\gamma = 2$ and a pupil of 250 mm, a dioptric design can accommodate up to 6 lenses with average overall transmission, or 8 lenses made of very low-absorption materials before a catadioptric design becomes preferable on throughput grounds alone.

It is important to note the limitations of this model. The assumption that obstruction is field- and wavelength-independent is optimistic: in practice, since the detector is not located at the pupil plane and has finite depth, the additional obstruction increases away from the optical axis and from the central wavelength. The model is therefore conservative in the sense that it underestimates the number of lenses a dioptric design can accommodate off-axis and off-wavelength. Nevertheless, the on-axis, central-wavelength comparison remains informative for design filtering purposes.

From this section, it can be concluded that transmission plays a central role in filtering viable designs. For detectors larger than 60mm, it is far preferable to use a dioptric design. For detectors with a side length of 60 mm, however, there are three distinct cases: for the 4k-15µm detector, given its aperture and étendue, it is worth exploring both types of design, but with greater interest in the dioptric design since the pupil of its catadioptric version will likely be noticeably larger (290 mm and 325 mm respectively for the dioptric and the catadioptric). In the case of an excessively fast camera, as with the 6k-10µm detector, since the estimated pupil diameter for the catadioptric version is slightly smaller than its dioptric counterpart, it would be preferable to explore only catadioptric designs. For the 6k-10µm detector with binning, the catadioptric pupil is estimated at 210 mm, 25 mm larger than the 185 mm dioptric pupil. Since this allows more than 10 lenses in the dioptric design before throughput parity is reached, the focus should be on the dioptric design only. The conclusions drawn here are summarised in Sec.~\ref{sec:designs}.

It should also be noted that the total throughput of a design with a large étendue per spectrograph will be penalised not only by the number of lenses required, but also by the need for stronger asphericisations and additional corrector elements to maintain image quality over the wider field. In this sense, the étendue-throughput coupling reinforces the conclusion from Sec.~\ref{sec:etendue}: simpler, more numerous spectrographs are preferable from a throughput standpoint, with each additional lens introducing $\sim$1.4\% loss assuming negligible absorption.

\subsubsection*{Detector noise budget and technology readiness}

While the CMOS market is rapidly evolving, the CCD market remains a niche with very few vendors, creating a risk that large quantities of CCDs may not be available or may become prohibitively expensive within the next ten years. Since most of the design options considered in Tab.~\ref{tab:list} require binning, this is a major concern: CCD detectors offer on-chip binning with low readout noise ($\sim$2~$e^-$) and low dark current ($\sim$2~$e^-$/hour), while CMOS detectors are not capable of this and would add noise that could interfere with observations.

To determine whether scientific CMOS detectors could nonetheless be viable, we derive the maximum allowable detector noise such that its contribution remains below 15\% of the sky photon noise in a single 20-minute exposure. It is defined as the quadratic sum of readout and dark current noise. The sky flux floor under dark-time conditions at zenith is taken from the Paranal atmospheric model\cite{Paranal1,Paranal2}. Within the IFS wavelength range, the minimum value is reached at 370 nm, where the sky flux is $350\,\mathrm{ph\,s^{-1}\,m^{-2}\,\mu m^{-1}\,arcsec^{-2}}$. Assuming a 20-minute integration, a telescope collecting area of 100~m$^2$, a total throughput of 25\%, and a spatial sampling of 0.25", the maximum detector noise such that it represents a fraction F of the total noise budget is given by:

\begin{equation}
N_\mathrm{det} = \sqrt{\frac{F \, p_\mathrm{sky}}{1 - F}},
\label{eq:det_noise}
\end{equation}

\noindent where $F = 15\%$ is the allowed noise fraction. Using the spectral sampling of a 90 mm detector design (0.48~$\mathring{\mathrm{A}}$ and 0.64~$\mathring{\mathrm{A}}$ for the blue and red channels respectively) we obtain sky fluxes ($p_\mathrm{sky}$) of 18.9 and 25.2 photons. The resulting noise requirements are $N_\mathrm{det} = 1.8\,e^-$ and $2.1\,e^-$ for the blue and red channels, assuming 2-pixel no-noise on-chip binning. In the absence of such binning, the spatial sampling effectively halves to 0.125", and the requirements tighten to $0.9\,e^-$ and $1.1\,e^-$ respectively. For 3-pixel no-noise on-chip binning, the corresponding requirements are $1.5\,e^-$ and $1.7\,e^-$; without on-chip binning, they drop to $0.5\,e^-$ and $0.6\,e^-$.

These figures have a direct bearing on the design trade-off. The requirement of sub-electron noise without on-chip binning is extremely demanding and unlikely to be met by current scientific CMOS technology across a large-format sensor. This means that design options with high binning ratios are effectively contingent on either CCD detectors or a significant advance in CMOS noise performance. The feasibility of on-chip no-noise binning in CMOS therefore represents a critical technology uncertainty that could significantly alter the ranking of design options, and should be monitored closely as the detector market evolves.

\subsection{Engineering and feasibility constraints}

\subsubsection*{Opto-mechanical volume: toy-model}

To compare the physical footprint of the different design options and assess whether they can be accommodated within the IFS instrument room, we construct a paraxial toy model that estimates the volume of the rectangular cuboid enclosing each spectrograph going from the entrance slit to the detector, and including optics and mechanics. The model is deliberately simple: its purpose is to provide order-of-magnitude estimates and establish relative comparisons between options, not to predict the volume of a specific optical design. The full parametric derivation and calibration against MUSE, BlueMUSE, 4MOST, and MOONS are given in Appendix~\ref{app:volume}. The adopted parameter values are $\tau_1 = 0.75$, $\tau_2 = 1$, $\tau_3 = 2$, $\tau_4 = 2.5$ for dioptric cameras, and $\tau_1 = 0.75$, $\tau_2 = 2.6$, $\tau_3 = 2$, $\tau_4 = 2.65$ for catadioptric cameras, with fixed parameters $\alpha = 10^\circ$, $N_\mathrm{col} = 3.6$, $m_\mathrm{opt} = 1.1$, and $m_\mathrm{meca} = 2.6$. Numerical volume estimates for all design options are given in Tab.~\ref{tab:volumes} in Appendix~\ref{app:volume}.

\begin{figure} [H]
\begin{center}
\begin{tabular}{c} 
\includegraphics[height=7.5cm]{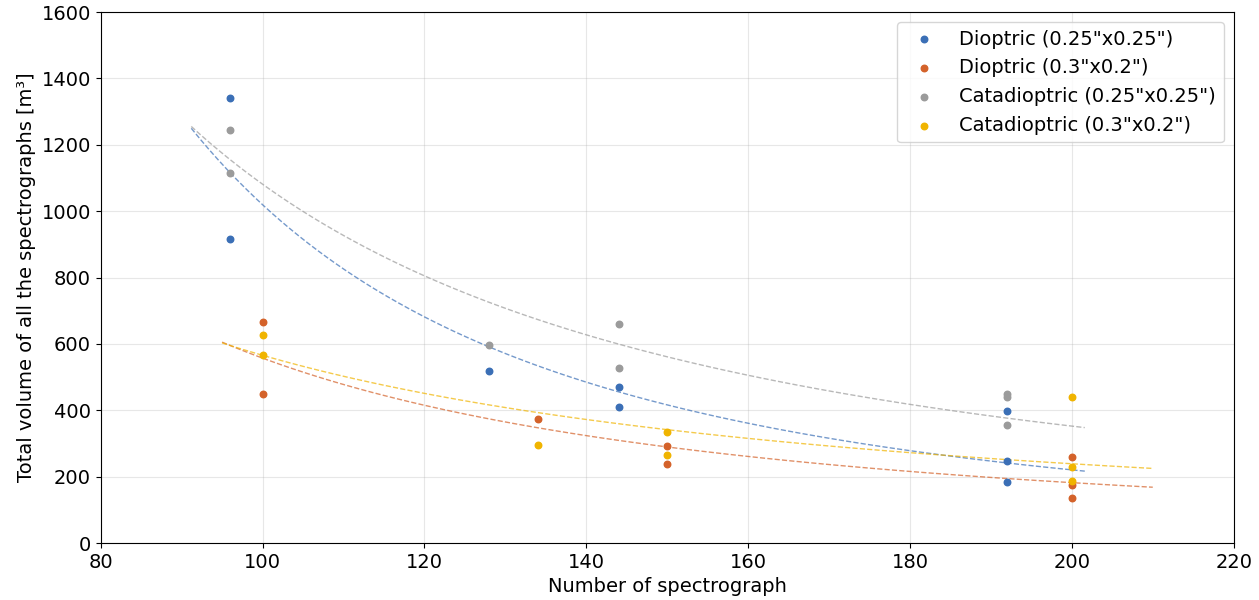}
\end{tabular}
\end{center}
\caption[SC] 
{ \label{fig:volumes} 
Total volume of the spectrographs as a function of the number of spectrographs, grouped by camera design and spatial sampling, with power-law fits.}
\end{figure}

A striking and counterintuitive result emerges from these estimates: design options with the smallest étendue per spectrograph --- and therefore the largest number of units --- also have the smallest total volume. This holds across all camera types and sampling options considered, as shown in Fig.~\ref{fig:volumes}. The reason is that the volume of a single spectrograph grows faster with its geometric extent than the total number of units decreases: more numerous, simpler spectrographs actually take up less space in aggregate.

\begin{figure} [H]
\begin{center}
\begin{tabular}{c} 
\includegraphics[height=8cm]{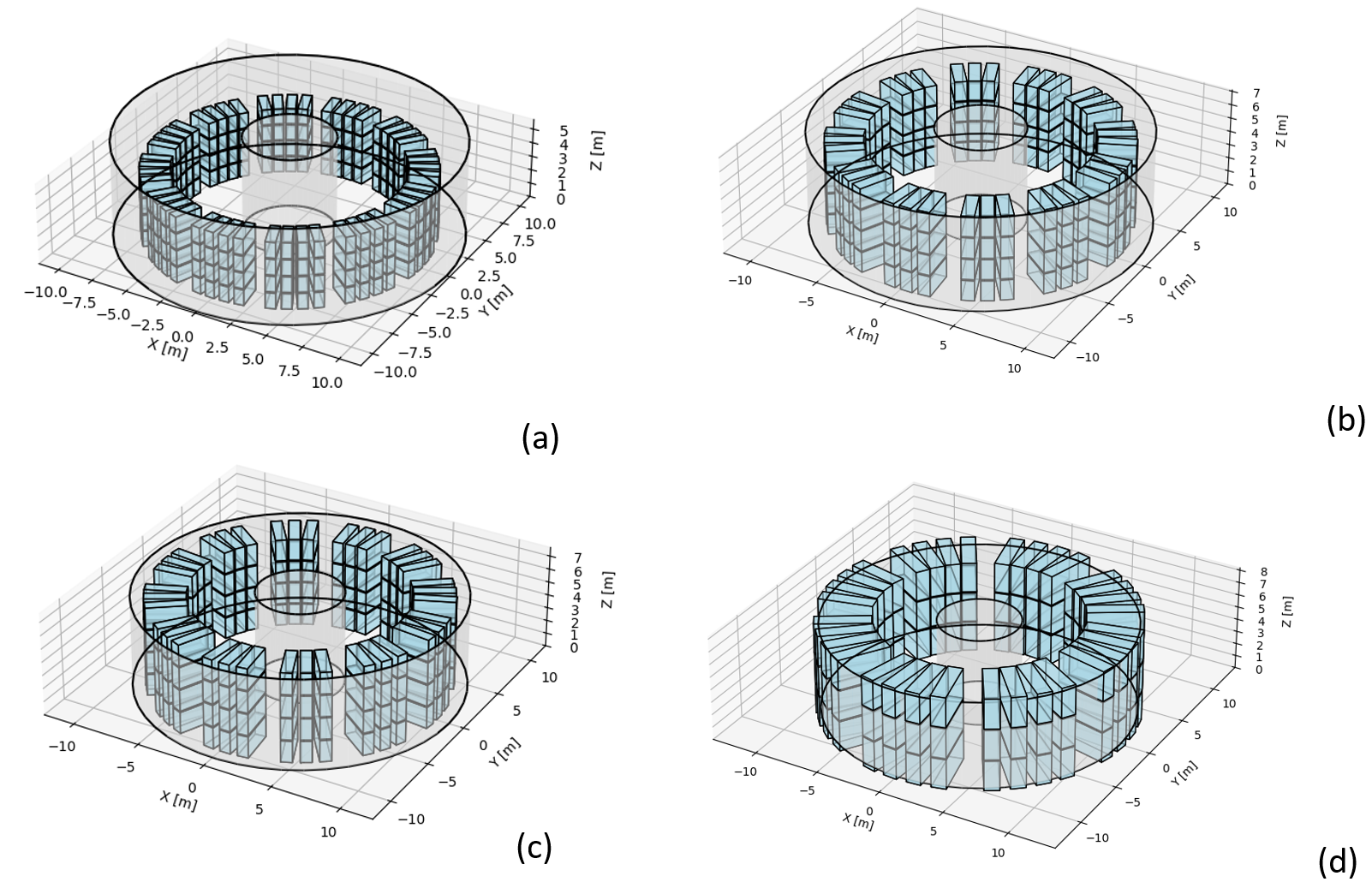}
\end{tabular}
\end{center}
\caption[SC] 
{ \label{fig:IFSroom} 
Estimated volume occupied by the spectrographs within the current IFS room for four representative dioptric design options: (a) 6k-10µm with 2$\times$1 binning, (b) 8k-10µm with 2$\times$1 binning, (c) 4k-15µm, and (d) 6k-15µm. The inner and outer cylinders are walls.}
\end{figure}

Fig.~\ref{fig:IFSroom} illustrates this concretely by showing the spectrograph layout within the IFS room for four representative options. Designs filling approximately 10\% of the room volume represent the practical upper limit for comfortable operation: beyond this, the spectrographs either reach the ceiling, or leave insufficient clearance for cryogenic systems, electronics, cabling, and personnel access. Options filling 15--20\% already reach the ceiling even at three spectrographs high, and those at 35\% filling would require the room to be substantially enlarged.

\subsubsection*{Practical MAIT (Manufacturing, Assembly, Integration, Testing) constraints}

Beyond the quantitative metrics above, several practical considerations further constrain the design space. Detector accessibility is particularly important in a catadioptric design, where the detector sits in the middle of the incident beam. This makes access for alignment, maintenance, and cryogenic integration significantly more complex than in a dioptric design. 

Designs requiring lens diameters in the 250--400 mm range present another practical constraint: very few glass types are available in such dimensions, and those that are (most notably CaF$_2$) carry costs that can make a design economically prohibitive.

%% file: 4-Draft_optical_designs.tex
\section{DRAFT OPTICAL DESIGNS COMPARISONS}
\label{sec:designs}

Building on the metrics and filtering criteria established in Sec.~\ref{sec:metrics}, this section carries out the actual design trade-off. We first use the accumulated constraints to reduce the full design space to a manageable set of viable options, then present draft optical designs for each, and finally compare them across throughput, volume, cost, and environmental impact.

\subsection{Design space and down-selection}
\label{sec:downselect}

\subsubsection*{Filtering on camera speed} 

The camera f-number is the single most consequential parameter in the early filtering of design options, as it simultaneously drives the detector and spectrograph error budgets, and the optical design complexity. As established in Sec.~\ref{sec:Perf}, cameras faster than f/1.35 push the detector error budget beyond what has been demonstrated on MUSE, and simultaneously tighten the spectrograph image quality budget to the point where only highly complex designs can comply. We therefore adopt f/1.35 as a practical lower bound on acceptable camera speed. 

Applying this threshold directly eliminates two detector formats. The 6k-10~µm detector without binning requires a camera at f/0.69, far beyond this limit, and is removed from consideration on those grounds alone. The 6k-15~µm detector without binning requires a camera at f/1.03 and carries an extremely large geometric étendue, making it both optically demanding and practically infeasible; it is also eliminated. The 4k-15~µm detector also requires f/1.03, but its relatively smaller étendue makes the resulting design more tractable. Hence, even though its f-number is beyond f/1.35, it is retained for now to at least consolidate the conclusions of the trade-off. 

\subsubsection*{Filtering on detector size and camera type} 

The throughput analysis of Sec.~\ref{sec:Perf} established that, for detectors larger than 60 mm, the additional central obstruction of a catadioptric design becomes prohibitive at any realistic pupil size and anamorphism ratio. Only detectors with a maximum of 60 mm side length are therefore compatible with a catadioptric architecture. Within this subset, the 6k-10~µm detector with 2$\times$1 binning is further excluded from the catadioptric exploration: its camera is slow enough that a dioptric design is the preferable option on both throughput and complexity grounds. 

\subsubsection*{Viable design options} 

The filtering steps above yield the seven viable combinations listed in Tab.~\ref{tab:ViableDesigns}. While all seven are technically feasible, they are not of equal interest. Several additional observations are relevant to their relative ranking:

\begin{itemize}
\item The 4k-15~µm option is the only one that requires reducing the spectral sampling to 2.0-2.1 pixels (rather than 2.4) to meet the spectral resolution requirement, leaving a much smaller margin for adjustment.
\item From a volume standpoint, only the options requiring 192 spectrographs fit comfortably within the current IFS room; designs with 144 spectrographs are marginal and would likely require the room to be enlarged.
\item On throughput, the ranking broadly follows the ordering by design simplicity, since fewer optical surfaces means less cumulative loss.
\end{itemize}

\begin{table}[H]
\caption{Viable combinations of sensors and camera designs after down-selection.} 
\label{tab:ViableDesigns}
\begin{center}       
\begin{tabular}{|>{\centering\arraybackslash}m{2.0cm}|>{\centering\arraybackslash}m{1.9cm}|>{\centering\arraybackslash}m{1.6cm}|>{\centering\arraybackslash}m{2.2cm}|>{\centering\arraybackslash}m{2cm}|>{\centering\arraybackslash}m{2.4cm}|>{\centering\arraybackslash}m{1.4cm}|} 
\hline
\rule[-1ex]{0pt}{3.5ex}  Detector & Camera design & Camera f/ & Number of spectrographs & Used spaxels per spectrograph & Estimated pupil diameter [mm] & Viable design ? \\
\hline
\rule[-1ex]{0pt}{3.5ex}  \multirow{2}{*}{4k-15µm} & \vphantom{\shortstack{X\\X}}Dioptric & 1.03 & 144 & 3600 & 290 & \textcolor{green!50!black}{\textbf{Yes}} \\
\cline{2-7}
\rule[-1ex]{0pt}{3.5ex}   & \vphantom{\shortstack{X\\X}}Catadioptric & 1.03 & 144 & 3600 & 325 & \textcolor{green!50!black}{\textbf{Yes}} \\
\hline
\rule[-1ex]{0pt}{3.5ex}  \multirow{2}{*}{6k-10µm} & \vphantom{\shortstack{X\\X}}Dioptric & 0.69 & 96 & 5400 & 540 & \textcolor{red}{\textbf{No}} \\
\cline{2-7}
\rule[-1ex]{0pt}{3.5ex}   & \vphantom{\shortstack{X\\X}}Catadioptric & 0.69 & 96 & 5400 & 495 & \textcolor{red}{\textbf{No}} \\
\hline
\rule[-1ex]{0pt}{3.5ex}  \multirow{2}{*}{6k-15µm} & \vphantom{\shortstack{X\\X}}Dioptric & 1.03 & 96 & 5400 & 415 & \textcolor{red}{\textbf{No}} \\
\cline{2-7}
\rule[-1ex]{0pt}{3.5ex}   & \vphantom{\shortstack{X\\X}}Catadioptric & 1.03 & 96 & 5400 & 460 & \textcolor{red}{\textbf{No}} \\
\hline
\rule[-1ex]{0pt}{3.5ex}  \multirow{2}{*}{\shortstack{6k-10µm\\(2x1 bin.)}} & \vphantom{\shortstack{X\\X}}Dioptric & 1.37 & 192 & 2700 & 185 & \textcolor{green!50!black}{\textbf{Yes}} \\
\cline{2-7}
\rule[-1ex]{0pt}{3.5ex}   & \vphantom{\shortstack{X\\X}}Catadioptric & 1.37 & 192 & 2700 & 210 & \textcolor{red}{\textbf{No}} \\
\hline
\rule[-1ex]{0pt}{3.5ex}  \multirow{2}{*}{\shortstack{8k-10µm\\(2x1 bin.)}} & \vphantom{\shortstack{X\\X}}Dioptric & 1.37 & 144 & 3600 & 240 & \textcolor{green!50!black}{\textbf{Yes}} \\
\cline{2-7}
\rule[-1ex]{0pt}{3.5ex}   & \vphantom{\shortstack{X\\X}}Catadioptric & 1.37 & 144 & 3600 & 260 & \textcolor{red}{\textbf{No}} \\
\hline
\rule[-1ex]{0pt}{3.5ex}  \multirow{2}{*}{\shortstack{9k-10µm\\(2x1 bin.)}} & \vphantom{\shortstack{X\\X}}Dioptric & 1.37 & 128 & 4050 & 270 & \textcolor{green!50!black}{\textbf{Yes}} \\
\cline{2-7}
\rule[-1ex]{0pt}{3.5ex}   & \vphantom{\shortstack{X\\X}}Catadioptric & 1.37 & 128 & 4050 & 280 & \textcolor{red}{\textbf{No}} \\
\hline
\rule[-1ex]{0pt}{3.5ex}  \multirow{2}{*}{\shortstack{9k-10µm\\(3x1 bin.)}} & \vphantom{\shortstack{X\\X}}Dioptric & 1.71 & 192 & 2700 & 200 & \textcolor{green!50!black}{\textbf{Yes}} \\
\cline{2-7}
\rule[-1ex]{0pt}{3.5ex}   & \vphantom{\shortstack{X\\X}}Catadioptric & 1.71 & 192 & 2700 & 205 & \textcolor{red}{\textbf{No}} \\
\hline
\rule[-1ex]{0pt}{3.5ex}  \multirow{2}{*}{\shortstack{6k-15µm\\(2x1 bin.)}} & \vphantom{\shortstack{X\\X}}Dioptric & 1.37 & 192 & 2700 & 140 & \textcolor{green!50!black}{\textbf{Yes}} \\
\cline{2-7}
\rule[-1ex]{0pt}{3.5ex}   & \vphantom{\shortstack{X\\X}}Catadioptric & 1.37 & 192 & 2700 & 195 & \textcolor{red}{\textbf{No}} \\
\hline
\end{tabular}
\end{center}
\end{table}

\subsection{Draft optical designs}
\label{sec:draft}

Draft optical designs have been produced for each viable option, and their main characteristics are summarised in Tab.~\ref{tab:draft}. Each design is assigned a letter from A to G. Where variants with curved detectors exist, a subscript distinguishes them: $f$ denotes a flat detector, $c$ a cylindrically curved one, and $t$ a toroidally curved one.

All designs assume two spectral arms: a single-arm design covering 370--930~nm in one pass would require a VPHG with unacceptably non-uniform diffraction efficiency, while three arms would add cost and complexity without scientific justification. The dichroic is therefore inserted in a collimated space shared by both arms.\cite{IFS_spectros}

\begin{table}[H]
\caption{Main characteristics of the draft designs.} 
\label{tab:draft}
\begin{center}       
\begin{tabular}{|>{\centering\arraybackslash}m{1.7cm}|>{\centering\arraybackslash}m{1.3cm}|>{\centering\arraybackslash}m{1.9cm}|>{\centering\arraybackslash}m{1.4cm}|>{\centering\arraybackslash}m{1.6cm}|>{\centering\arraybackslash}m{1.8cm}|>{\centering\arraybackslash}m{1.8cm}|>{\centering\arraybackslash}m{2.0cm}|} 
\hline
\rule[-1ex]{0pt}{3.5ex}  Detector & Name & Camera design & Detector curvature & Real pupil diameter [mm] & Lowest throughput in both arms & Highest throughput in both arms & Approximate volume per spectrograph [$m^3$] \\
\hline
\rule[-1ex]{0pt}{3.5ex}  \multirow{2}{*}{4k-15µm} & \vphantom{\shortstack{X\\X}}A & Dioptric & Flat & 270 & 81.4\% & 82.3\% & 3.28\\
\cline{2-8}
\rule[-1ex]{0pt}{3.5ex}   & \vphantom{\shortstack{X\\X}}B & Catadioptric & Flat & 340 & 80.7\% & 84.6\% & 4.90 \\
\hline
\rule[-1ex]{0pt}{3.5ex}  \shortstack{9k-10µm\\(2x1 bin.)} & C & Dioptric & Flat & 256 & 80.6\% & 81.6\% & 4.10\\
\hline
\rule[-1ex]{0pt}{3.5ex}  \shortstack{8k-10µm\\(2x1 bin.)} & D & Dioptric & Flat & 233 & 82.3\% & 82.8\% & 2.48\\
\hline
\rule[-1ex]{0pt}{3.5ex}  \shortstack{6k-10µm\\(2x1 bin.)} & $E_f$ & Dioptric & Flat & 200 & 86.5\% & 86.7\% & 1.14\\
\hline
\rule[-1ex]{0pt}{3.5ex}  \shortstack{6k-10µm\\(2x1 bin.)} & $E_c$ \& $E_t$ & Dioptric & Curved & 203 & 89.0\% & 89.2\% & 1.05\\
\hline
\rule[-1ex]{0pt}{3.5ex}  \shortstack{9k-10µm\\(3x1 bin.)} & $F_f$ & Dioptric & Flat & 175 & 86.5\% & 86.7\% & 1.64\\
\hline
\rule[-1ex]{0pt}{3.5ex}  \shortstack{9k-10µm\\(3x1 bin.)} & $F_c$ \& $F_t$ & Dioptric & Curved & 175 & 89.0\% & 89.2\% & 1.14\\
\hline
\rule[-1ex]{0pt}{3.5ex}  \shortstack{6k-15µm\\(2x1 bin.)} & $G_f$ & Dioptric & Flat & 150 & 86.6\% & 86.7\% & 0.97\\
\hline
\rule[-1ex]{0pt}{3.5ex}  \shortstack{6k-15µm\\(2x1 bin.)} & $G_c$ \& $G_t$ & Dioptric & Curved & 150 & 89.0\% & 89.2\% & 0.88\\
\hline
\end{tabular}
\end{center}
\end{table}

\subsubsection*{Collimator design}

A fully dioptric collimator, as used in MUSE and BlueMUSE, would be problematic for WST for three reasons. First, the collimator length would be impractical: at f/7 with a 200 mm pupil, the collimated space would span approximately 2800 mm. Second, the large slit length at this f-ratio (around 300 mm for 3000 spaxels) would require very large input lenses. Third, correcting the differential aberrations between the two spectral arms in the collimated space before the dichroic would likely require 4--6 lenses. 

\begin{figure} [H]
\begin{center}
\begin{tabular}{c} 
\includegraphics[height=4cm]{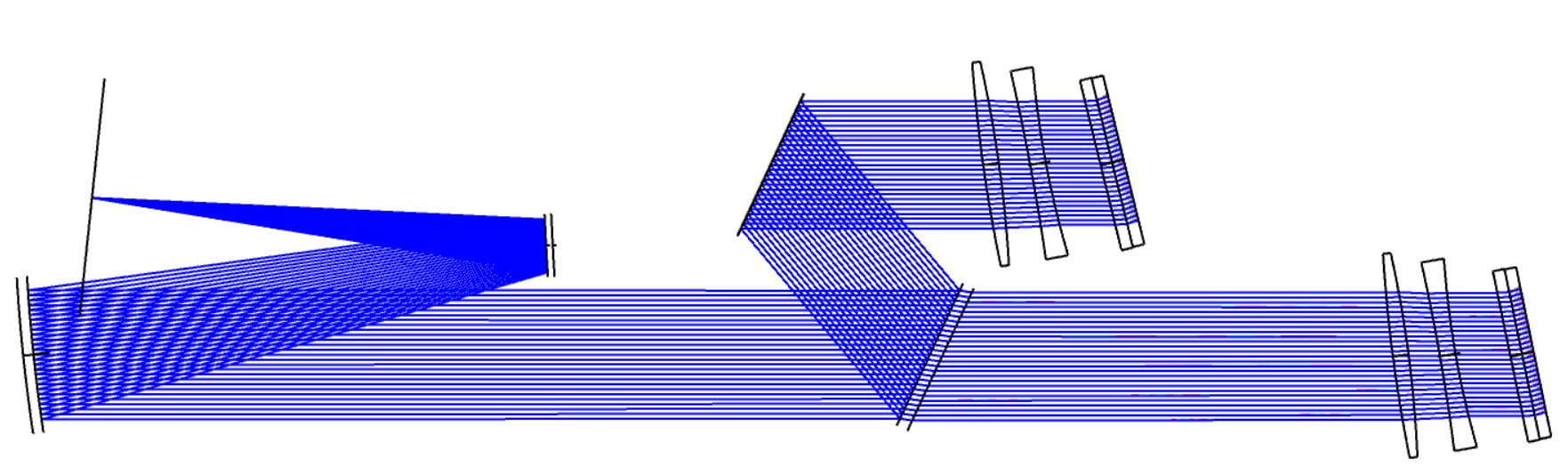}
\end{tabular}
\end{center}
\caption[SC] 
{ \label{fig:Collimator} 
Layout of the selected reflective collimator design, consisting of a spherical mirror, a conic mirror, a dichroic, and an afocal airspaced silica doublet in each spectral arm.}
\end{figure}

Instead, we designed a reflective collimator inspired by multi-object spectrograph collimator architectures, converging on the layout shown in Fig.~\ref{fig:Collimator}. It consists of just two mirrors: a sphere followed by a pure conic, followed by an afocal airspaced silica doublet in each arm, placed after the dichroic. The doublet lenses are "simple" elements (no wedge or freeform terms) mounted with a small tilt and decentering with respect to the optical axis, which makes the alignment insensitive to the rotational clocking of individual components. Compared to a dioptric collimator, this design requires fewer optical surfaces, a shorter mechanical envelope, and is better suited to the insertion of a dichroic in a well-controlled collimated space.

\subsubsection*{Camera designs}

\begin{figure} [H]
\begin{center}
\begin{tabular}{c} 
\includegraphics[height=7.6cm]{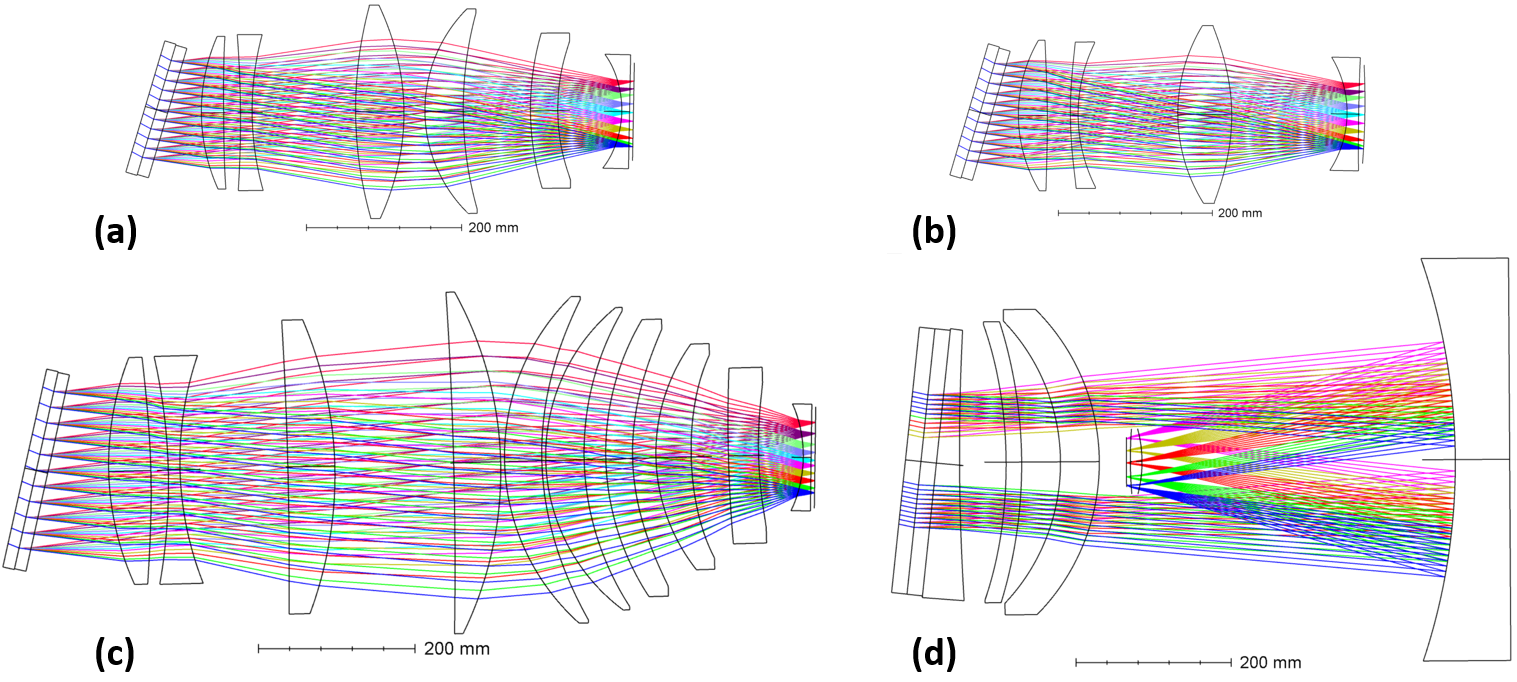}
\end{tabular}
\end{center}
\caption[SC] 
{ \label{fig:Camera} 
Examples of draft camera optical designs at the same scale: (a) 6k-15µm with 2$\times$1 bin. and flat detector, (b) 6k-15µm with 2$\times$1 bin. and curved detector, (c) 9k-10µm with 2$\times$1 bin., (d) 4k-15µm catadioptric.}
\end{figure}

To maximise throughput, particularly at the blue end of the spectrum, and to limit the variety of glass types, the camera designs use only three materials: Silica, CaF$_2$, and PBL35Y. The first two are multi-sourceable; the third is sourced from Ohara with an equivalent available in Schott’s catalog under reference LF5HTi. S-FPL53 or other special glasses could be substituted for CaF2 if required, though this would reduce the spectrograph's average transmission by $\sim$5\% due to internal absorption. According to the manufacturers we contacted, it is also not particularly easier to work with. Examples of draft camera layouts are shown in Fig.~\ref{fig:Camera}.

The more demanding camera designs clearly require more challenging optical components: larger elements, stronger lens bending, and larger aspheric departures, as confirmed by the risk analysis carried out by manufacturers (Tab.~\ref{tab:RISK}). Conversely, the simpler designs benefit considerably from relaxed constraints, and in particular from the use of curved detectors\cite{HighlyCurved,Olaf,TorCurved} : both cylindrically and toroidally curved sensors allow up to two lenses to be removed per camera, as seen in the transition from design (a) to design (b) in Fig.~\ref{fig:Camera}. This simplification increases the optical throughput by approximately 2.5\% per arm. However, curved detectors are complex and risky to produce; there will likely be a decrease in yield and an increase in cost as well as in error budgets if they are chosen.

\begin{table}[H]
\caption{Preliminary internal risk analysis provided by the manufacturers.} 
\label{tab:RISK}
\begin{center}       
\begin{tabular}{|c|c|c|c|} 
\hline
\rule[-1ex]{0pt}{3.5ex}  Risk & Description & Design with a flat detector & Design with a curved detector  \\
\hline
\rule[-1ex]{0pt}{3.5ex}  A & Material Availability (technical) & None & None   \\
\hline
\rule[-1ex]{0pt}{3.5ex}  B & Material Availability (production) & Low & Low  \\
\hline
\rule[-1ex]{0pt}{3.5ex}  C & Component manufacturing & None & None \\
\hline
\rule[-1ex]{0pt}{3.5ex}  D & Optical coatings & None & None \\
\hline
\rule[-1ex]{0pt}{3.5ex}  E & Integration and tests & None & None \\
\hline
\rule[-1ex]{0pt}{3.5ex}  F & Sensor & None & Unknown \\
\hline
\end{tabular}
\end{center}
\end{table}

The throughput values quoted in Tab.~\ref{tab:draft} exclude the dichroic, grating diffraction efficiency, and detector quantum efficiency, and assume a coating transmission of 99.3\% per optical surface. As expected, the more complex designs are strongly constrained in the throughput they can achieve, since each additional surface required to meet the image quality budget reduces transmission. For reference, the BlueMUSE spectrograph transmission budget is 84\% (excluding the same terms and similar coating efficiency); only the simplest WST design options exceed this value. This further reinforces the case for simple, numerous spectrographs.

\begin{figure} [H]
\begin{center}
\begin{tabular}{c} 
\includegraphics[height=6.5cm]{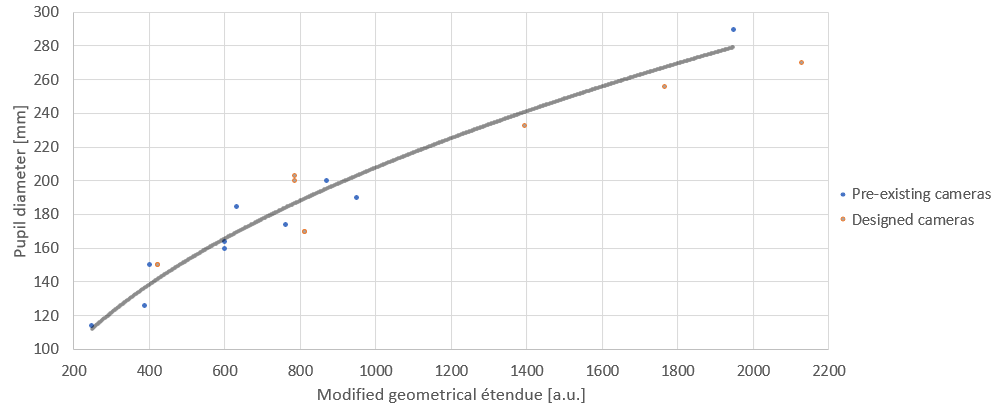}
\end{tabular}
\end{center}
\caption[SC] 
{ \label{fig:RealPupil} 
Pupil diameters of the designed cameras overlaid on the empirical fit from Fig.~\ref{fig:D_pup_diop_app}.}
\end{figure}

As a validation of the pupil diameter scaling relations from Sec.~\ref{sec:pupil}, Fig.~\ref{fig:RealPupil} shows the designed pupil diameters overlaid on the empirical fit: the agreement is good, with a maximum discrepancy of 8.8\% and a mean of 5.6\%, consistent with the expected accuracy of the model for designs within or near the calibration range.

\subsection{Opto-mechanical volume comparison}

The volume estimates for the draft designs confirm the trend identified by the toy model: the volume of a single spectrograph grows rapidly with its aperture and field size, faster than the reduction in unit count. The conclusions are unchanged: designs with 192 spectrographs are the only ones that fit comfortably within the current IFS room, while designs with 128 or 144 spectrographs would require the room to be enlarged to varying degrees. The catadioptric option (design B) has an estimated volume so large that a major extension of the IFS room would be required.

\newpage

\subsection{Cost analysis}
\label{sec: cost}

\subsubsection*{Objective and structure}

The cost analysis serves two purposes: to compare the design options against each other and identify which are likely to be significantly more expensive, and to provide an order-of-magnitude estimate of the total cost of the selected architecture. For the comparison, we use parametric equations that are calibrated against the available reference data. For the absolute estimate, we use cost data provided by manufacturers.
To simplify the cost estimation, the overall budget is divided into two components: the cost of the spectrograph optomechanics (including materials and MAIT) and the cost of the detectors and image slicers. The cost model proceeds in four steps:

\begin{itemize}
    \item First, computation of the Theoretical First Unit (TFU) cost.

    \item Next, application of a risk factor to account for the Technology Readiness Level (TRL).

    \item Then, incorporation of the serial effect, reflecting productivity gains as more units are produced.

    \item Finally, adjustment for inflation to January 2025 euros.
\end{itemize}

\subsubsection*{Theoretical First Unit (TFU) cost}

The TFU represents the MAIT cost of producing the first spectrograph unit, including all fixed engineering costs for the development of the complete series. The intention is to compute it via the simplest possible expression that can be calibrated using data from relevant past projects. Drawing on Meinel's power-law scaling of telescope cost with aperture \cite{Meinel}, we assume that the TFU scales with the relative cost of the lenses' material and the optics diameter raised to a given power:

\begin{equation}
TFU = K\sum_{i \leq n_{surf}}\alpha_iD_{i}^A,
\label{eq:TFU}
\end{equation}

\noindent where $D_i$ is the diameter of surface $i$, $K$ and $A$ are constants calibrated against three financial proposals. $\alpha_i$ is a parameter corresponding to the relative cost of the material of the surface $i$. The TFU is composed of the fixed engineering costs and the recurring costs, and based on the same financial proposals, we assume that recurring costs represent 30\% of the TFU ($\tau = 0.30$), giving:

\begin{equation}
C_{MAIT,fix} = (1-\tau)TFU = (1-\tau)K\sum_{i \leq n_{surf}}\alpha_iD_{i}^A,
\label{eq:MAITFIX}
\end{equation}

\begin{equation}
C_{MAIT,rec} = \tau TFU = \tau K\sum_{i \leq n_{surf}}\alpha_iD_{i}^A,
\label{eq:MAITREC}
\end{equation}

Raw material costs are modelled with a separate power law, similar to the one presented for the MAIT part,

\begin{equation}
C_{mat} = K'\sum_{i \leq n_{surf}}\alpha_iD_{i}^{A'},
\label{eq:mat}
\end{equation}

\noindent where K' and A' are constants calibrated against the same financial proposals. We acknowledge that this model is far from perfect and would clearly need to be more refined with more data points and a more complex modelling. The $\alpha_i$ are obtained by maximising the fits, and in practice this parameter is not used for the MAIT, and $\alpha_{CaF2}$ = 6.7 while the rest is set to 1 for the raw material cost. The resulting calibration constants are:

\begin{equation}
K = 6.84\times10^{-5} \ a.u., \quad A = 2.32, \quad K' = 2.97\times10^{-7} \ a.u., \quad A' = 2.83
\label{eq:calib}
\end{equation}

\subsubsection*{Risk factor}

Following the approach of Miller, Keesee, and Jilla\cite{Meinel}, which focuses on space projects, we propose introducing a risk factor (RF) to account for the Technology Readiness Level (TRL) associated with the different design options. The risk factor adds between $<$5\% (TRL 8, fully operational) and $>$25\% (TRL 1--2, conceptual) to the estimated cost.

\subsubsection*{Learning curve}

The serial effect is estimated using a learning curve. This type of curve is used in many industries to take account of the impact on the costs of productivity gains when many units of a given product are produced\cite{Learn}. We choose to use the DeJong model because, in addition to accounting for the serial effect $\phi$, it also captures the fraction ($1-M$) of production where learning cannot occur, typically the part carried out by machines:

\begin{equation}
C = C_1\times\bigg((1-M)n_s + M\sum_{i=1}^{n_s}i^B\bigg), \quad \text{with} \quad B=\frac{\mathrm{log}(\phi)}{\mathrm{log}(2)}
\label{eq:LEARNING}
\end{equation}

\noindent where C is the total recurring cost of the $n_s$ units and $C_1$ the recurring cost of the very first one. $\phi$ is the learning rate, the factor by which the marginal cost is multiplied each time the total production doubles and $M$ is the fraction of $C_1$ subject to the serial effect. 

Finally, the overall cost for the spectrographs writes:

\begin{equation}
C = C_{MAIT,fix} + C_{MAIT,rec}\times\bigg((1-M)n_s + M\sum_{i=1}^{n_s}i^B\bigg) + C_{mat}n_s
\label{eq:C_spectros}
\end{equation}

Based on the polishing time data from VLT, Gemini, and JWST mirror fabrication (Stahl \& Allison 2020), we anticipate a learning rate in the range $\phi \in [0.75, 0.85]$. This is consistent with standard aerospace industry values.

\begin{figure} [H]
\begin{center}
\begin{tabular}{c} 
\includegraphics[height=5.5cm]{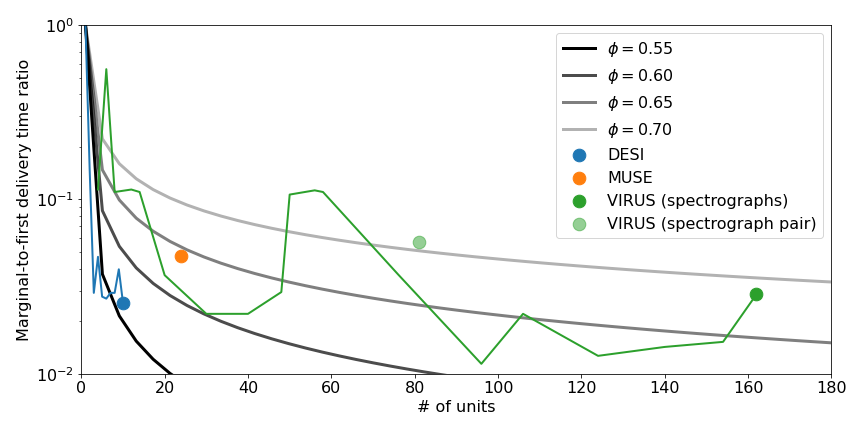}
\end{tabular}
\end{center}
\caption[SC] 
{ \label{fig:LR3} 
Marginal-to-first delivery time ratio for DESI\cite{DESI} , MUSE\cite{MUSE}, and VIRUS\cite{VIRUS} spectrographs, with learning curve fits.}
\end{figure}

Analysis of actual delivery time data from DESI ($n = 10$), MUSE ($n = 24$), and VIRUS ($n = 156$) suggests somewhat higher productivity gains ($\phi \in [0.55, 0.70]$), as shown in Fig.~\ref{fig:LR3}. In the case of VIRUS, each optomechanical unit consists of a spectrograph pair; however, as shown, whether units are counted individually or as pairs has little effect on the estimated learning rate. 
This may reflect the higher proportion of manual labour in spectrograph MAIT compared to mirror fabrication, though the limited sample size makes it difficult to draw firm conclusions.
It is also possible that delivery dates are not a reliable enough proxy to allow us to estimate a learning rate. Indeed, they largely depend on external factors that do not directly affect costs.

Given the uncertainty, we adopt the conservative range $\phi \in [0.75, 0.85]$. Following informal consultation with manufacturers, $M$ is estimated at approximately 50\%, and we model it as $M \in [0.45, 0.55]$.

\subsubsection*{Detector cost}

Forecasting CMOS detector costs is inherently uncertain given the rapidly evolving market. Hence, for this estimation we conservatively assume a cost similar to those of scientific CCD sensors. We assume that it is independent of format or pixel pitch, with specifications similar to those of MUSE/BlueMUSE, which as we said in Sec.~\ref{sec:Perf} would likely not be true for very fast cameras. Combining these assumptions leads to a cost estimate that scales linearly with the total number of detectors. To simplify the estimation and because it remains unknown, we conservatively assume that there is no serial effect on the cost of the detectors. The equation used for this estimation is the following:

\begin{equation}
C_{det} = 2n_s\left(C_{det,0} + \frac{CC}{NDC} + CSC + \frac{AC}{NDA} + MC + ACC\right)
\label{eq:C_det}
\end{equation}

\noindent where the factor of 2 accounts for two arms per spectrograph, $C_{det,0}$ is the base detector cost set at 130 k€. \textbf{CC} corresponds to the controller cost, \textbf{NDC} is the number of detectors by controller, \textbf{CSC} is the cooling system cost, \textbf{AC} is the aggregator, \textbf{NDA} is the number of detectors by aggregator, \textbf{MC} is the mechanics cost, and \textbf{ACC} is the additional cost for the curvature.

\subsubsection*{Image slicer cost}

Concerning the image slicers, their cost may be a small fraction of the total cost, depending on the technology used to produce them. If they are polished like on MUSE, they will be very expensive, but if they are machined, their cost will massively drop. We choose to include them, and as for the detectors, we assume that the total cost scales linearly with the total number of image slicers and number of slices per slicer:

\begin{equation}
C_{IS} = C_{IS,unit}n_{slices}n_s
\label{eq:C_IS}
\end{equation}

\noindent with $C_\mathrm{IS,unit} =$ $\sim8$~k\euro/slice, consistent with a recent slicer's cost per slice, and $n_{slices}$ is the estimated number of slices need for each design. 

This estimation is very pessimistic and assumes a manufacturing process incompatible with the construction timeline of the WST facility. With approximately 150-200 image slicers, machining is the only viable option if production is not to take 20 years. However, without analogous cost estimates for such slicers, we cannot develop our own estimate. Thus, this portion of the total cost is very penalising for design options with the largest number of IFUs. This assumption undermines the final conclusion (which leans towards numerous simple spectrographs), making a cheaper per-unit image slicer advantageous anyway.

\subsubsection*{Total cost and Monte Carlo uncertainty}

The toy model is thus constructed by combining the terms proposed above for the cost of spectrographs, detectors, and image slicers, yielding the following equation:

\begin{equation}
C_{tot} = C_{MAIT,fix} + C_{MAIT,rec}\times\bigg((1-M)n_s + M\sum_{i=1}^{n_s}i^B\bigg) + C_{mat}n_s + 2C_{det,unit}n_s + C_{IS,unit}n_{slices}n_s
\label{eq:C_tot}
\end{equation}

This equation can be rewritten using the parameters and constants described above as:

\begin{align}
C_{tot} =
&(1-\tau)K\sum_{i \leq n_{surf}}D_{i}^A + \tau K\sum_{i \leq n_{surf}}D_{i}^A\bigg((1-M)n_s + M\sum_{i=1}^{n_s}i^B\bigg) + K'\sum_{i \leq n_{surf}}\alpha_iD_{i}^{A'}n_s \nonumber\\
&+ 2C_{det,unit}n_s + C_{IS,unit}n_{slices}n_s,
\label{eq:C_tot2}
\end{align}

\begin{figure} [H]
\begin{center}
\begin{tabular}{c} 
\includegraphics[height=5.0cm]{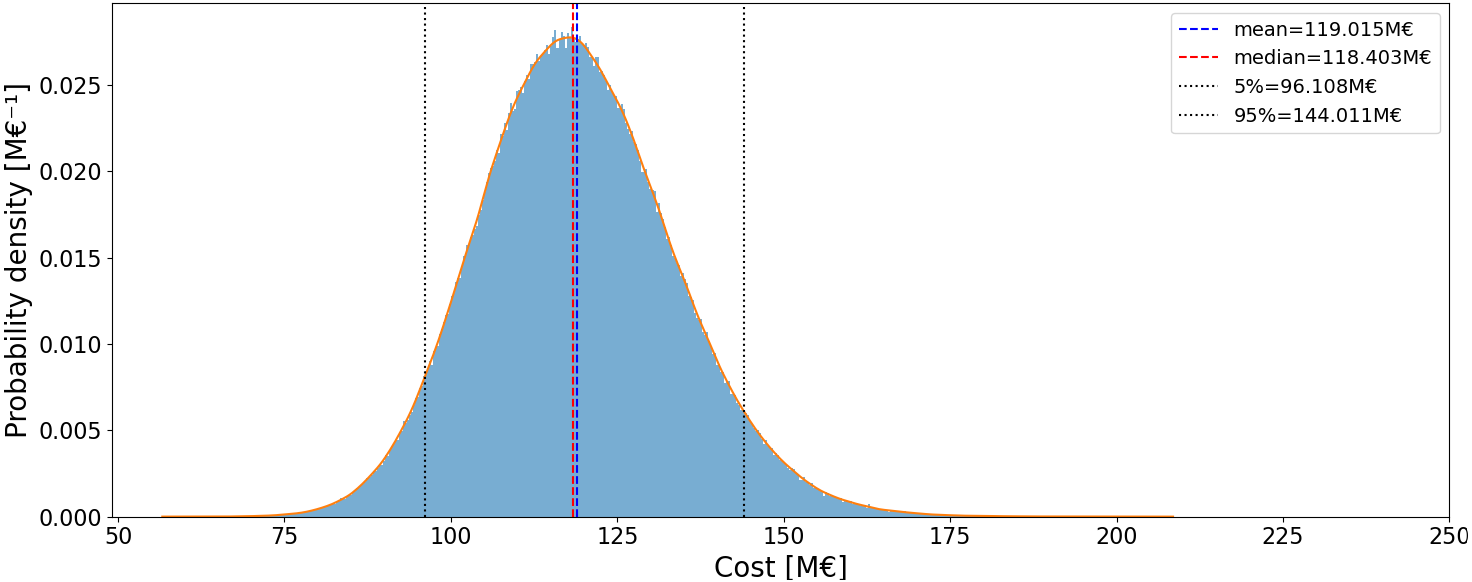}
\end{tabular}
\end{center}
\caption[SC] 
{ \label{fig:Cost_histo} 
Example cost distribution from the Monte Carlo simulation.}
\end{figure}

Using the values for $n_s$, $n_{surf}$ and $D_{i}$ obtained in the draft designs and this toy model, we can compute cost estimations for each design option still considered. Since $\tau$, $\phi$, and $M$ are not precisely known, they are sampled from their respective intervals via Monte Carlo simulation to quantify the impact of this arbitrary uncertainty on the total cost. The resulting distributions are well-behaved (Fig.~\ref{fig:Cost_histo}), and we report the 5\%, median, and 95\% quantiles for each design option as a central estimate with a confidence interval. More precisely, we anticipate a learning rate about 75-85\%, hence we assume $\phi\in[0.75,0.85]$; since $\tau$ has been observed to be close to 30\% in other projects, we presume $\tau\in[0.25,0.35]$; and as said previously we suppose that $M\in[0.45,0.55]$.

\subsubsection*{Cost estimations}

\begin{figure} [H]
\begin{center}
\begin{tabular}{c} 
\includegraphics[height=9.9cm]{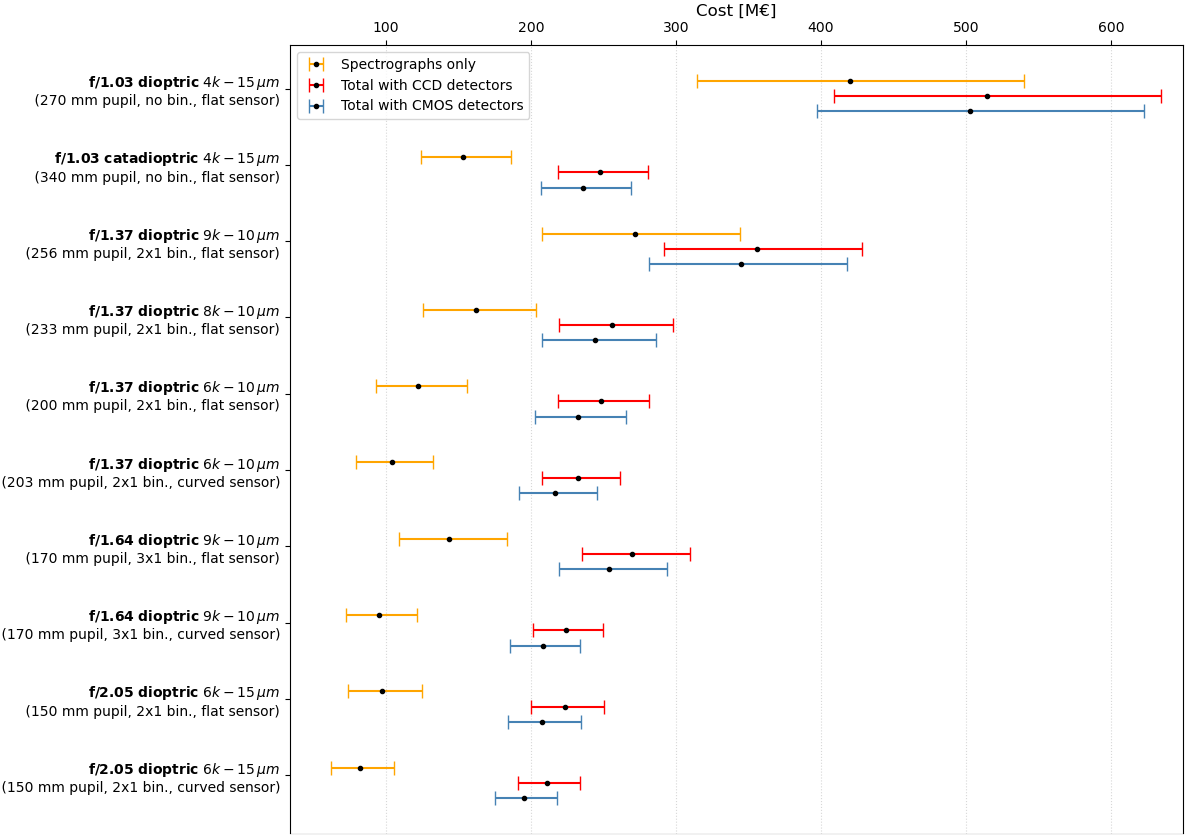}
\end{tabular}
\end{center}
\caption[SC] 
{ \label{fig:Cost_estim} 
Cost estimates from the toy model for all viable design options, showing spectrograph-only costs and total costs including detectors and image slicers, with 5\%--95\% confidence intervals.}
\end{figure}

The cost estimates for all viable design options are shown in Fig.~\ref{fig:Cost_estim}. Several clear trends emerge.

Complex designs, despite requiring fewer units, are consistently more expensive in total. The unit cost grows so steeply with optics diameter that it more than offsets the savings from building fewer spectrographs. Conversely, simpler designs cost less in total even after accounting for the larger number of units, because the learning curve effect reduces the per-unit cost significantly over a long series and amortizes the engineering costs over many units. For a fixed number of spectrographs, a slower camera with a larger detector seems preferable to a faster one with a smaller detector, as the weighted sum of the diameters is slightly smaller. The catadioptric option is in-between; it is neither the most nor the least expensive, but note that this toy model does not take into account additional costs that may be necessary for a catadioptric camera.

Curved detectors consistently reduce the total cost across all design options because removing two lenses per arm outweighs the additional curving cost. This holds provided the per-detector curving cost remains low and provided the curved design does not require larger optics to achieve similar image quality. The second condition is non-trivial, especially for higher-étendue designs.

Note that this toy model is an estimation tool, not a precise cost predictor, as it does not try to represent the real manufacturing scheme. It should therefore be interpreted as indicative rather than predictive, and as a guide for selecting a reference architecture. Comparison with the manufacturer ROM estimates shows the model underestimates spectrograph costs by a factor of 1.1--2.3 (Tab.~\ref{tab:ROM_spectro}), likely because of the assumed values for $\tau$, $\phi$ and $M$: they almost certainly didn't factor in any serial effect. But note that even without accounting for serial effects, the conclusions drawn from the toy model remain unchanged and favor numerous simple designs. For image slicers (Tab.~\ref{tab:ROM_slicers}), the two estimates are already in good agreement, given the limited number of assumptions involved. The manufacturer's current estimates also imply manufacturing lead times of 10--20 years at existing industrial capacity, without accounting for any expansion of production capability or economies of scale.

\begin{table}[ht]
\caption{ROM cost estimates from the manufacturer compared to the toy model, for selected design options.} 
\label{tab:ROM_spectro}
\begin{center}       
\begin{tabular}{|c|c|c|c|} 
\hline
\rule[-1ex]{0pt}{3.5ex}  Design option & \shortstack{4k-15µm\\(flat, no binning)} & \shortstack{6k-10µm\\(flat, 2x1 binning)} & \shortstack{6k-10µm\\(curved, 2x1 binning)}  \\
\hline
\rule[-1ex]{0pt}{3.5ex}  Spectrographs cost only [M€] & 460 & 260 & 240   \\
\hline
\rule[-1ex]{0pt}{3.5ex}  Toy model estimation [M€] & 420 & 122 & 104  \\
\hline
\end{tabular}
\end{center}
\end{table}

\begin{table}[ht]
\caption{ROM costs given by the manufacturer for the image slicers.} 
\label{tab:ROM_slicers}
\begin{center}       
\begin{tabular}{|c|c|c|} 
\hline
\rule[-1ex]{0pt}{3.5ex}  Image slicer & \shortstack{Design A\\(30 slices, 192 units)}  & \shortstack{Design B\\(30 slices, 192 units)}   \\
\hline
\rule[-1ex]{0pt}{3.5ex}  Slicers cost only [M€] & 57 & 60   \\
\hline
\rule[-1ex]{0pt}{3.5ex}  Toy model estimation [M€] & 46 & 46  \\
\hline
\end{tabular}
\end{center}
\end{table}

\subsubsection*{Integral Field Units cost trends}

This trade-off study seeks to determine whether it is more advantageous to build a large number of simple spectrographs or a smaller number of more complex units. The toy model shows that the answer, from a cost perspective, is not straightforward. In particular, two competing trends appear in the model:

\begin{align}
C_{tot} =
&(1-\tau)K\sum_{i \leq {\color{red}n_{surf}}}{\color{red}D_{i}}^A + \tau K\sum_{i \leq {\color{red}n_{surf}}}{\color{red}D_{i}}^A\bigg((1-M){\color{green!50!black}n_s} + M\sum_{i=1}^{{\color{green!50!black}n_s}}i^B\bigg) + K'\sum_{i \leq {\color{red}n_{surf}}}\alpha_i{\color{red}D_{i}}^{A'}{\color{green!50!black}n_s} \nonumber\\
&+ 2C_{det,unit}{\color{green!50!black}n_s} + C_{IS,unit}n_{slices}{\color{green!50!black}n_s},
\label{eq:C_trend}
\end{align}

The parameters shown in red and green both contribute to an increase in cost when they increase. However, these two sets of parameters evolve in opposite directions: when the red parameters increase, the green ones generally decrease, and vice versa. This reflects the fact that increasing the étendue per spectrograph leads to more optical surfaces and a larger pupil diameter, while reducing the total number of spectrographs required.

This equation highlights several features of the toy model, in particular the existence of an optimal étendue per spectrograph. Two trends can be identified. First, when the étendue per spectrograph is small, the number of spectrographs is large, and the total cost is mainly driven by the detectors and image slicers. Second, as the étendue per spectrograph increases, the total cost rises. Although one might expect that reducing the number of spectrographs by making each unit more complex would lower the overall cost, the opposite is observed here: the increase in unit cost outweighs the savings from having fewer units. Interestingly, as seen in Fig.~\ref{fig:Cost_estim}, this model predicts that at a given étendue, a slower camera is cheaper even though its detector is larger. This characteristic comes from the slightly stronger increase in lens size with aperture than with field size, or said otherwise, for a given number of spectrographs, $\sum_{i \leq n_{surf}}D_{i}^A$ is lower when the camera is slower.

\subsection{Environmental impact assessment}

The carbon footprint of the instrument provides a complementary criterion to the economic and performance metrics, and supports more sustainable design choices from the earliest stages of development. We evaluate the CO$_2$-equivalent emissions associated with the construction phase and 20 years of operation, using SimaPro (version 10.2.0.1) with the ReCiPe 2016 Midpoint default assessment method. The functional unit is the construction and operation of the IFS for one continuous year, including its cooling system and readout electronics.

The construction footprint covers the raw materials and manufacturing of the optical elements, the spectrographs, their cooling system and readout electronics. The housing of the spectrographs is excluded, as the designs are not yet sufficiently advanced to provide housing geometries for all options. For the CCD system, we assume a Linear Pulse Tube cryocooler (Thales 9310) with ESO New Generation Controller II \cite{Carbon1} readout electronics. For the CMOS system, we assume a Peltier-plus-CO$_2$ cooling system with FPGA-based readout. 

The operational footprint is computed solely based on the power consumption of the various systems, using the current Chilean grid energy mix. Data processing power is not included in the current assessment.

\begin{figure} [H]
\begin{center}
\begin{tabular}{c} 
\includegraphics[height=6cm]{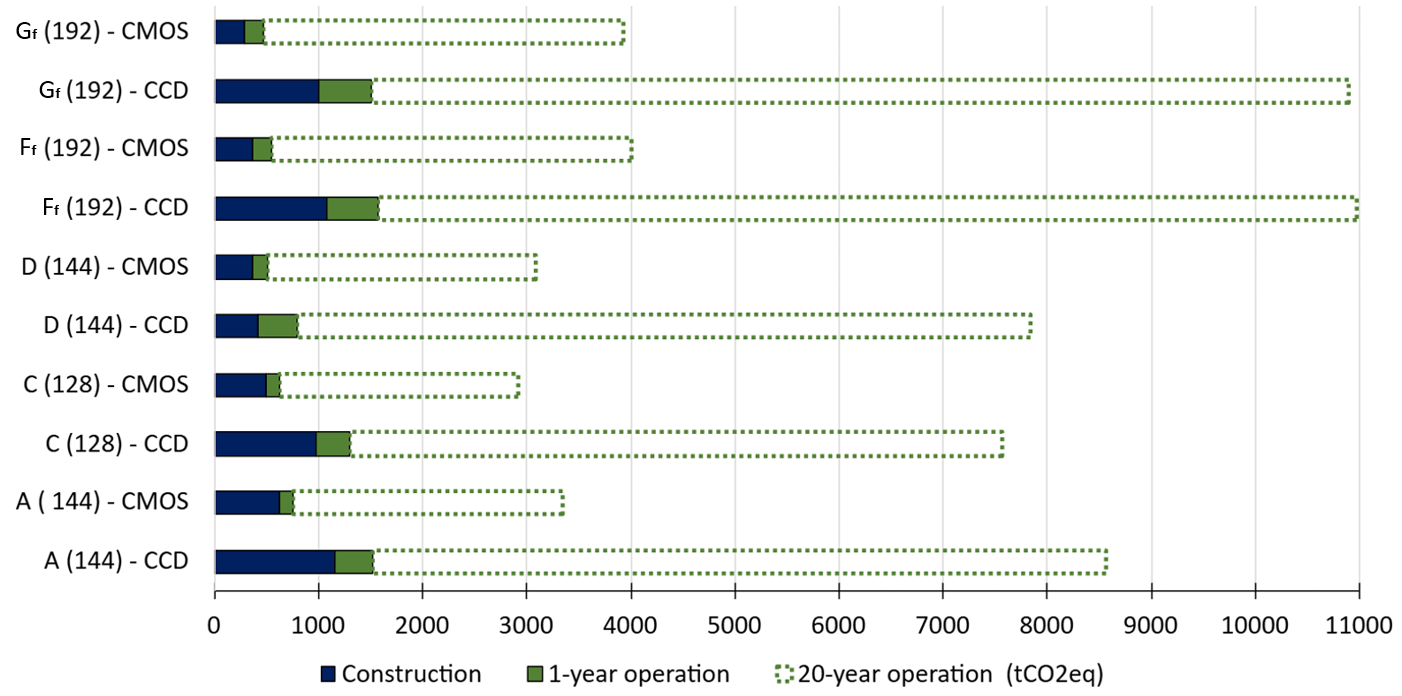}
\end{tabular}
\end{center}
\caption[SC] 
{ \label{fig:Carbon} 
Carbon footprint of the construction (blue) and operation (green) for different design options.}
\end{figure}

Fig.~\ref{fig:Carbon} shows the carbon footprint estimated using the Global Warming impact category of the SimaPro results for the construction (in blue), the one-year operation (in green) and the 20-year operation (in dashed green) for ten different design options, five different optical designs and spectrograph configurations $A$, $C$, $D$, $F_f$, and $G_f$ and CCD and CMOS for each. The number of spectrographs for each option is indicated inside brackets. For all designs, the operational footprint eventually exceeds that of construction after a few years. We note here that the enclosures of the spectrographs and optical bench are not included in the analysis due to a lack of data, but they might have a non-negligible impact on construction due to the amount of aluminium or stainless steel required, resulting in associated metalworking energy\cite{Carbon2}.

The 20-year operational footprint is dominated by detector cooling and readout power consumption, and grows with unit count (Fig.~\ref{fig:Carbon}). Across the range of viable designs, the 20-year footprint increases from 2.6~Mt CO$_2$eq (design A, 144 spectrographs, CMOS) to 3.5~Mt CO$_2$eq (design $F_f$, 192 spectrographs, CMOS), a ~25\% increase driven by the difference in unit count. Choosing design C (128 spectrographs) instead of $F_f$ would decrease the carbon footprint by 35\%. Focusing on design $F_f$, a switch from CCD to CMOS would reduce the carbon footprint by 63\%. Importantly, technology choice and unit count interact: the operational footprint of design A with CCDs (7.04~Mt) exceeds that of designs with 192 spectrographs and CMOS (3.45~Mt), illustrating that a lower unit count does not automatically translate into a lower environmental impact if the detector technology and cooling systems are less efficient. Overall, CMOS can operate at higher temperatures, enabling a $\mathrm{CO_2}$-based cooling system that can be shared among the telescope's three instruments. This results in a smaller construction footprint and lower energy consumption for CMOS-based options. We note that the readout electronics also play a role here. FPGAs usually have lower energy consumption than NGC II systems. Consequently, replacing NGC II with FPGAs in the options that use CCDs could reduce the carbon footprint, but it would not change the ranking of the designs based on carbon footprint.

%% file: 5-Trade-off_matrix.tex
\section{TRADE-OFF MATRIX}
\label{sec:tradeoff}

Following the evaluation of performance metrics, engineering constraints, cost, and environmental impact presented in Sections~\ref{sec:metrics} and~\ref{sec:designs}, a global comparison of the remaining design options is performed using a trade-off matrix. This approach provides a structured and transparent framework for ranking the candidate architectures by combining heterogeneous criteria into a single decision metric.

\subsection{Structure and scoring methodology}

The trade-off matrix is constructed from the seven criteria identified throughout this study: image quality, throughput, spectral properties, system size, cost and risk, detector accessibility, and environmental impact. Each design is assigned a score between 1 (least favourable) and 5 (most favourable) for each criterion.

For quantitative metrics, scores are assigned by linear normalisation between the minimum and maximum values obtained across all designs. The best-performing design receives a score of 5 and the worst a score of 1, with intermediate values interpolated linearly. For criteria where lower values are preferable (cost, volume, and environmental impact) the scaling is inverted accordingly. For qualitative criteria such as detector accessibility, scores are assigned on the basis of engineering judgment, accounting for integration complexity, maintenance access, and operational constraints as discussed in Sec.~\ref{sec:metrics}. These are mapped onto the same 1--5 scale to ensure consistency with the quantitative criteria.

\subsection{Weighting scheme}

\begin{figure}[H]
\begin{center}
\begin{tabular}{c}
\includegraphics[height=5.0cm]{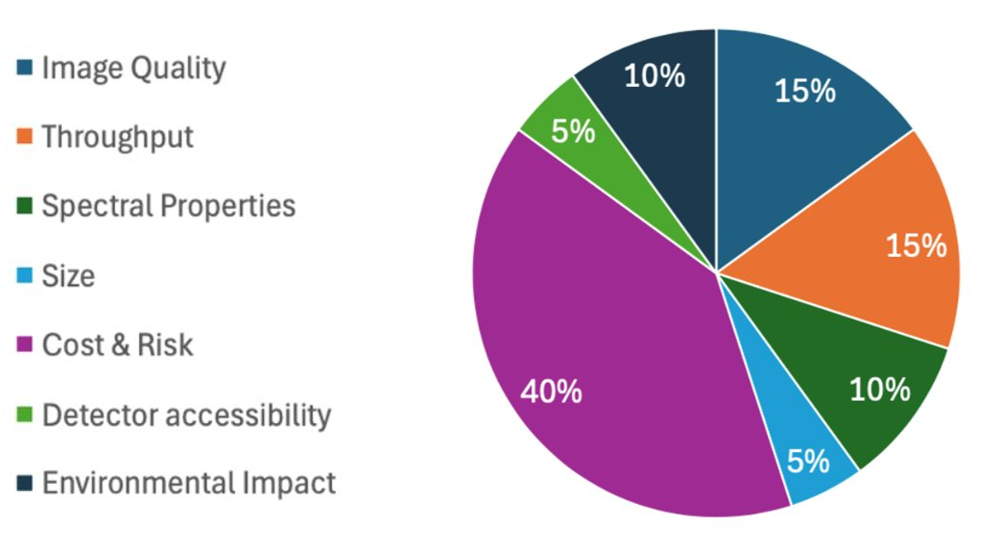}
\end{tabular}
\end{center}
\caption[SC]
{ \label{fig:PieChart}
Weight distribution for the seven criteria used in the trade-off matrix.}
\end{figure}

The attribution of weights is a top-level decision that can significantly influence the final ranking. Depending on project priorities, one may favour scientific performance, cost containment, or operational simplicity. We adopt the joint weighting scheme illustrated in Fig.~\ref{fig:PieChart}, which reflects the conclusions drawn throughout this analysis.

Performance-related criteria account for 40\% of the total weight: image quality and throughput are each assigned 15\%, reflecting their direct impact on scientific return, while spectral properties carry 10\%, since the requirements are met by most designs with comfortable margin. Cost and risk are assigned the largest single weight, at 40\%, consistent with the strong sensitivity of the total budget to optical complexity and unit count identified in Sec.~\ref{sec: cost}. Environmental impact is included at 10\%, reflecting its growing relevance as a design criterion for large facilities. System size and detector accessibility each carry 5\%, capturing integration and operational constraints that are important but secondary to performance and cost. The weighted total score is:

\begin{equation}
S = \sum_{i} w_i \cdot s_i,
\label{eq:weighting}
\end{equation}

\noindent where $s_i$ is the score for criterion $i$ and $w_i$ its corresponding weight.

\subsection{Results}

The resulting trade-off matrix is presented in Tab.~\ref{tab:TradeOffMatrix}, ordered by decreasing total score and discussed below.

\begin{table}[H]
\caption{Trade-off matrix for the evaluated design options, ordered by decreasing weighted total score.}
\label{tab:TradeOffMatrix}
\begin{center}
\begin{tabular}{|>{\centering\arraybackslash}m{1.3cm}|
                 >{\centering\arraybackslash}m{1.2cm}|
                 >{\centering\arraybackslash}m{1.8cm}|
                 >{\centering\arraybackslash}m{1.6cm}|
                 >{\centering\arraybackslash}m{1.1cm}|
                 >{\centering\arraybackslash}m{1.2cm}|
                 >{\centering\arraybackslash}m{1.9cm}|
                 >{\centering\arraybackslash}m{1.3cm}|
                 >{\centering\arraybackslash}m{1.6cm}|}
\hline
\rule[-1ex]{0pt}{3.5ex} Design 
                        & Error budget (15\%) 
                        & Throughput (15\%) 
                        & Spectral properties (10\%) 
                        & Volume (5\%) 
                        & Cost \& Risk (40\%) 
                        & Det. accessibility (5\%) 
                        & Env. impact (10\%) 
                        & \textbf{Weighted total} \\
\hline
\rule[-1ex]{0pt}{3.5ex} {\color{green!80!black}$\mathrm{F_t}$} & 4.2 & 5.0 & 4.1 & 3.0 & 4.8 & 5.0 & 3.3 & \textbf{4.45} \\
\hline
\rule[-1ex]{0pt}{3.5ex} {\color{green!80!black}$\mathrm{G_t}$} & 4.3 & 5.0 & 3.0 & 3.2 & 5.0 & 5.0 & 3.3 & \textbf{4.43} \\
\hline
\rule[-1ex]{0pt}{3.5ex} {\color{green!80!black}$\mathrm{G_c}$} & 4.2 & 5.0 & 3.0 & 3.2 & 5.0 & 5.0 & 3.3 & \textbf{4.42} \\
\hline
\rule[-1ex]{0pt}{3.5ex} {\color{green!80!black}$\mathrm{F_c}$} & 3.9 & 5.0 & 4.1 & 3.0 & 4.8 & 5.0 & 3.3 & \textbf{4.40} \\
\hline
\rule[-1ex]{0pt}{3.5ex} {\color{orange!99!black}$\mathrm{G_f}$} & 4.3 & 3.7 & 3.0 & 3.1 & 4.9 & 5.0 & 3.3 & \textbf{4.18} \\
\hline
\rule[-1ex]{0pt}{3.5ex} {\color{orange!99!black}$\mathrm{E_t}$} & 2.9 & 5.0 & 3.0 & 3.1 & 4.7 & 5.0 & 3.3 & \textbf{4.09} \\
\hline
\rule[-1ex]{0pt}{3.5ex} {\color{orange!99!black}$\mathrm{E_c}$} & 2.7 & 5.0 & 3.0 & 3.1 & 4.7 & 5.0 & 3.3 & \textbf{4.06} \\
\hline
\rule[-1ex]{0pt}{3.5ex} {\color{orange!99!black}$\mathrm{F_f}$} & 4.3 & 3.7 & 4.1 & 2.6 & 4.2 & 5.0 & 3.3 & \textbf{4.00} \\
\hline
\rule[-1ex]{0pt}{3.5ex} {\color{red}$\mathrm{E_f}$} & 2.7 & 3.7 & 3.0 & 3.0 & 4.5 & 5.0 & 3.3 & \textbf{3.79} \\
\hline
\rule[-1ex]{0pt}{3.5ex} {\color{red}D} & 2.7 & 1.6 & 3.7 & 2.4 & 4.5 & 5.0 & 4.6 & \textbf{3.67} \\
\hline
\rule[-1ex]{0pt}{3.5ex} {\color{red}B} & 1.0 & 1.6 & 1.0 & 1.0 & 4.8 & 1.0 & 4.6 & \textbf{3.12} \\
\hline
\rule[-1ex]{0pt}{3.5ex} {\color{red}C} & 2.7 & 1.0 & 4.1 & 1.7 & 2.5 & 5.0 & 5.0 & \textbf{2.80} \\
\hline
\rule[-1ex]{0pt}{3.5ex} {\color{red}A} & 1.2 & 1.4 & 1.0 & 2.0 & 1.0 & 5.0 & 4.6 & \textbf{1.69} \\
\hline
\end{tabular}
\end{center}
\end{table}

The results reveal a clear and consistent ranking structure that reinforces the conclusions drawn in the individual metric analyses.

\textbf{Top tier (Score $\geq$ 4.40): designs {\color{green!80!black}$\mathrm{F_t}$}, {\color{green!80!black}$\mathrm{G_t}$}, {\color{green!80!black}$\mathrm{G_c}$}, {\color{green!80!black}$\mathrm{F_c}$}:}

The four highest-ranked designs all correspond to architectures with 192 spectrographs and score strongly on cost and risk (4.7--5.0). They all achieve the maximum throughput score (5.0) owing to their cylindrical or toric detectors. Within this sub-group, the $\mathrm{F}$ designs (9k-10µm, 3$\times$1 binning: $\mathrm{F_t}$, $\mathrm{F_c}$) score higher on spectral properties (4.1 vs 3.0) than the $\mathrm{G}$ designs (6k-15µm, 2$\times$1 binning: $\mathrm{G_t}$, $\mathrm{G_c}$), while the $\mathrm{G}$ designs achieve the maximum cost score (5.0 vs 4.8 for the $\mathrm{F}$ group). Design $\mathrm{F_t}$ ranks first overall, as its spectral advantage marginally outweighs its cost disadvantage at the adopted weights. 

\newpage

\textbf{Middle tier (4.00 $\leq$ Score $\leq$ 4.18): designs {\color{orange!99!black}$\mathrm{G_f}$}, {\color{orange!99!black}$\mathrm{E_t}$}, {\color{orange!99!black}$\mathrm{E_c}$}, {\color{orange!99!black}$\mathrm{F_f}$}:}

At the top of this tier are designs $\mathrm{G_f}$ (4.18) and $\mathrm{E_t}$ (4.09): $\mathrm{G_f}$ compensates its flat-detector throughput penalty (3.7) through a strong cost score (4.9) and error budget (4.3), while $\mathrm{E_t}$ recovers full throughput via its toric detector but is penalised by a weaker error budget (2.9). $\mathrm{E_c}$ (4.06) shares the maximum throughput score with the curved top-tier designs but is held back by its weak error budget (2.7). $\mathrm{F_f}$ (4.00) benefits from a strong error budget (4.3) and spectral properties score (4.1), but its flat detector imposes a throughput penalty (3.7) and its cost score (4.2) falls noticeably below those of $\mathrm{F_c}$ and $\mathrm{F_t}$, relegating it out of the top tier despite its otherwise favourable profile.

\textbf{Lower tier (Score $\leq$ 3.79): designs {\color{red}$\mathrm{E_f}$}, {\color{red}$\mathrm{D}$}, {\color{red}$\mathrm{C}$}, {\color{red}$\mathrm{A}$}, {\color{red}$\mathrm{B}$}:}

$\mathrm{E_f}$ (3.79) ranks first in the lowest tier, combining the throughput penalty of a flat detector with the error budget penalty of the 6k-10µm option. The four other lowest-ranked designs are the only ones with less than 192 IFUs and are primarily penalised by low throughput and error budget scores. Design D (8k-10µm, 2$\times$1 binning) achieves a reasonable cost score (4.5) and scores well on environmental impact (4.6), but its throughput score (1.6) severely limits its total. Design C (9k-10µm, 2$\times$1 binning) achieves the best environmental score (5.0) and a strong spectral properties score (4.1), but its very low throughput (1.0) and poor cost score (2.5) pull its total down significantly. Design A (4k-15µm dioptric) ranks last overall, penalised across nearly every criterion: its fast f/1.03 camera drives the minimum cost score (1.0) and a poor error budget (1.2). Design B (4k-15µm catadioptric) scores reasonably on cost and risk (4.8), but its catadioptric design incurs the minimum detector accessibility score (1.0), the lowest volume score (1.0), and poor spectral properties (1.0) and throughput (1.6), collectively dragging its total to 3.12.

\subsection{Sensitivity and limitations}

The ranking presented above is robust at the extremes: the gap between the top tier (designs $F_t$--$F_c$) and the bottom tier (designs $E_f$--D) is large enough that plausible variations in the weighting scheme are unlikely to alter their relative positions, given the dominant role of the cost and risk criterion (40\%) in separating the tiers. Within the top tier, however, the scores of $F_t$, $G_t$, $G_c$, and $F_c$ are separated by 0.05 points or less, making their relative ranking sensitive to the precise weights attributed to spectral properties and cost. In particular, increasing the weight of spectral properties beyond 10\% would consolidate the advantage of $F_t$ and $F_c$, since the 9k-10µm detector achieves a noticeably higher spectral properties score (4.1 vs 3.0 for the $G$ group). Conversely, placing greater emphasis on cost would favour the $G$ designs, which achieve the maximum cost score (5.0 vs 4.8 for the $F$ group). The position of $G_f$ within the middle tier is moderately sensitive: its throughput score (3.7) is lower than the curved and toric designs (5.0), so any substantial downward reweighting of throughput could push it into the top tier. Within each detector family, curved and toric variants outperform their flat counterparts primarily through their throughput advantage (5.0 vs 3.7); their lead is therefore robust unless the throughput criterion is substantially downweighted. Only a very large increase in the risk associated with the curving process could bring a flat design above its curved equivalent.

It should also be noted that the cost and risk scores for curved-detector designs already incorporate the additional curving cost and the associated sensor risk, as discussed in Sec.~\ref{sec: cost}. Their advantage over flat counterparts in the ranking is therefore a net result, accounting for both the throughput gain from lens removal and the additional cost and risk of curved sensors. This trade-off matrix should nonetheless be regarded as a decision-support tool rather than a definitive ranking, and should be updated as detector technology matures and more precise cost estimates become available.

%% file: 6-Conlusion.tex
\newpage
\section{CONCLUSION}
\label{sec:conclusions}

This study has systematically explored the design space of the WST IFS spectrographs across a wide range of detector formats, camera architectures, and spatial sampling options. The analysis consistently points in the same direction: across nearly every metric considered (throughput, volume, cost, and construction carbon footprint) a larger number of simpler spectrographs is preferable to a smaller number of complex ones. This conclusion is not obvious a priori, and emerges from the steep scaling of unit cost and volume with lens diameter, which more than offsets the savings from building fewer units.

\subsubsection*{Why not using even slower cameras?}

One natural question raised by this analysis is whether pushing the camera focal ratio even slower than the designs studied here would be beneficial. The error budget analysis gives a clear lower bound: cameras faster than f/1.35 push the detector flatness requirement beyond what has been demonstrated on MUSE, and simultaneously tighten the spectrograph image quality budget to the point where only very complex designs can comply. The instrument therefore benefits from cameras as slow as possible, but this comes at a cost: for a fixed detector size, a slower camera requires more spectrographs. Beyond a certain point, an excessive number of units creates its own problems: a highly constrained production and integration schedule, and a large operational carbon footprint driven by the cooling and readout systems.

The natural resolution of this tension is to slow the cameras while simultaneously increasing the detector size, so that the number of spectrographs remains constant. A larger, slower camera also tends to have smaller optics than a faster camera with a smaller detector serving the same number of spaxels, which directly reduces unit cost and volume. However, larger detectors generally require more spatial binning, which introduces two further constraints. First, if on-chip no-noise binning cannot be achieved on CMOS sensors, as discussed in Sec.~\ref{sec:metrics}, then either CCD detectors will be required, or the designs with high binning ratios will have to be discarded. Second, increasing the binning ratio reduces the anamorphism ratio. Provided this ratio remains above 1, the design is unaffected; but if it falls below 1, the spectral axis of the camera becomes faster than the spatial axis, forcing a more demanding design than the spatial sampling alone would require, with corresponding increases in volume, cost, and tolerancing burden.

\subsubsection*{Selected design options}

The trade-off matrix of Sec.~\ref{sec:tradeoff} identifies three detector configurations as the most promising baseline options for the WST IFS, each available in flat and curved detector variants: the 6k-10µm with 2$\times$1 binning (designs $E_f$, $E_c$, $E_t$), the 9k-10µm with 3$\times$1 binning (designs $F_f$, $F_c$, $F_t$), and the 6k-15µm with 2$\times$1 binning (designs $G_f$, $G_c$, $G_t$). All require 192 spectrographs with a spectral sampling of 2.4 pixels, and their key characteristics are compared in Tab.~\ref{tab:Conclusion}.

The two families at the top of the ranking present a clear scientific trade-off. The 9k-10µm designs ($F_f$--$F_t$) offer substantially higher spectral resolution and represent the more scientifically ambitious option. However, their use is conditional on the availability of on-chip no-noise binning, since the 3$\times$1 binning required by this format cannot be performed without noise penalty on current CMOS sensors. Additionally, the anamorphism ratio of 0.8 places this design in the sub-unity regime, adding complexity to the camera optimisation. The 6k-15µm designs ($G_f$--$G_t$) avoid both of these issues: the 2$\times$1 binning requirement is less demanding, the anamorphism ratio is a more comfortable 1.2, and the total cost is approximately 10--50~M€ lower depending on the detector technology. These designs comfortably meet the current spectral resolution requirements, albeit with less margin than the 9k-10µm option. The E-family designs rank lower, primarily due to their faster spatial focal ratio (f/1.37) and higher associated costs. Across all three families, curved detector variants consistently outperform their flat counterparts on both throughput and total cost, and their advantage is robust in the trade-off matrix even after accounting for the additional curving cost and sensor risk.

\begin{table}[H]
\caption{Comparison of the three selected design options (flat and curved detector cases).}
\label{tab:Conclusion}
\begin{center}
\begin{tabular}{|>{\centering\arraybackslash}m{4.0cm}|>{\centering\arraybackslash}m{1.5cm}|>{\centering\arraybackslash}m{1.5cm}|>{\centering\arraybackslash}m{1.5cm}|>{\centering\arraybackslash}m{1.5cm}|>{\centering\arraybackslash}m{1.5cm}|>{\centering\arraybackslash}m{1.5cm}|}
\hline
\rule[-1ex]{0pt}{6ex} \multirow{2}{*}{Design option} & 
\multicolumn{2}{c|}{6k-10µm (2$\times$1 bin.)} & 
\multicolumn{2}{c|}{9k-10µm (3$\times$1 bin.)} & 
\multicolumn{2}{c|}{6k-15µm (2$\times$1 bin.)} \\
\cline{2-7}
\rule[-1ex]{0pt}{3.5ex} & Flat & Curved & Flat & Curved & Flat & Curved \\
\hline
\rule[-1ex]{0pt}{3.5ex} Design name & $E_f$ & $E_c$ \& $E_t$ & $F_f$ & $F_c$ \& $F_t$ & $G_f$ & $G_c$ \& $G_t$ \\
\hline
\rule[-1ex]{0pt}{3.5ex} Number of spectrographs & \multicolumn{6}{c|}{192} \\
\hline
\rule[-1ex]{0pt}{3.5ex} Spectral sampling [pix] & \multicolumn{6}{c|}{2.4} \\
\hline
\rule[-1ex]{0pt}{3.5ex} Camera spatial f/ & \multicolumn{2}{c|}{f/1.37} & \multicolumn{2}{c|}{f/2.06} & \multicolumn{2}{c|}{f/2.06} \\
\hline
\rule[-1ex]{0pt}{3.5ex} Camera spectral f/ & \multicolumn{2}{c|}{f/1.64} & \multicolumn{2}{c|}{f/1.65} & \multicolumn{2}{c|}{f/2.47} \\
\hline
\rule[-1ex]{0pt}{3.5ex} Anamorphism ratio & \multicolumn{2}{c|}{1.2} & \multicolumn{2}{c|}{0.8} & \multicolumn{2}{c|}{1.2} \\
\hline
\rule[-1ex]{0pt}{3.5ex} Average spectral resolution B/R & \multicolumn{2}{c|}{4040/4800} & \multicolumn{2}{c|}{6060/7200} & \multicolumn{2}{c|}{4040/4800} \\
\hline
\rule[-1ex]{0pt}{3.5ex} Lowest spectral resolution B/R & \multicolumn{2}{c|}{2960/3720} & \multicolumn{2}{c|}{4440/5580} & \multicolumn{2}{c|}{2960/3720} \\
\hline
\rule[-1ex]{0pt}{3.5ex} Total volume ratio & 1.0 & 0.92 & 1.44 & 1.0 & 0.85 & 0.77 \\
\hline
\rule[-1ex]{0pt}{3.5ex} Median cost [M€] (spectro. only) & 122 & 104 & 144 & 95 & 98 & 83 \\
\hline
\rule[-1ex]{0pt}{3.5ex} Median cost [M€] (with CCD) & 248 & 233 & 270 & 224 & 224 & 211 \\
\hline
\rule[-1ex]{0pt}{3.5ex} Median cost [M€] (with CMOS) & 232 & 217 & 254 & 208 & 208 & 195 \\
\hline
\end{tabular}
\end{center}
\end{table}

In summary, the 6k-15µm dioptric designs with curved detectors --- in particular with cylindrical curvatures, as they are less risky than toroidal --- emerge as the most prefered design options for the WST IFS spectrographs, offering the best balance of cost, throughput, volume, and design risk. Their counterpart with flat detectors reaches a lower score but is almost as promising and must, at the very least, be kept as a fallback solution if an architecture with curved detectors turns out to be unfeasible.

%% file: Appendix.tex
\newpage
\section{Derivations and additional tables}

\subsection{Pupil diameter scaling: calibration}
\label{app:pupil}

This appendix details the empirical calibration of the pupil diameter scaling relations used in Sec.~\ref{sec:pupil} (Eqs.~\ref{eq:Dpup_mod} and~\ref{eq:Dpup_cata}). It is provided for reproducibility; the trade-off analysis in Sec.~\ref{sec:designs} depends only on the fitted parameter values and the resulting estimates in Tab.~\ref{tab:Dpup}. Two constraints are imposed on the calibration sample to keep it representative of our regime of interest: only designs with pupil diameters above 100~mm are retained, and at least one arm must cover a spectral range comparable to the blue arm of WST (370--595~nm).

\subsubsection*{Dioptric calibration}

\begin{figure} [H]
\begin{center}
\begin{tabular}{c}
\includegraphics[height=6cm]{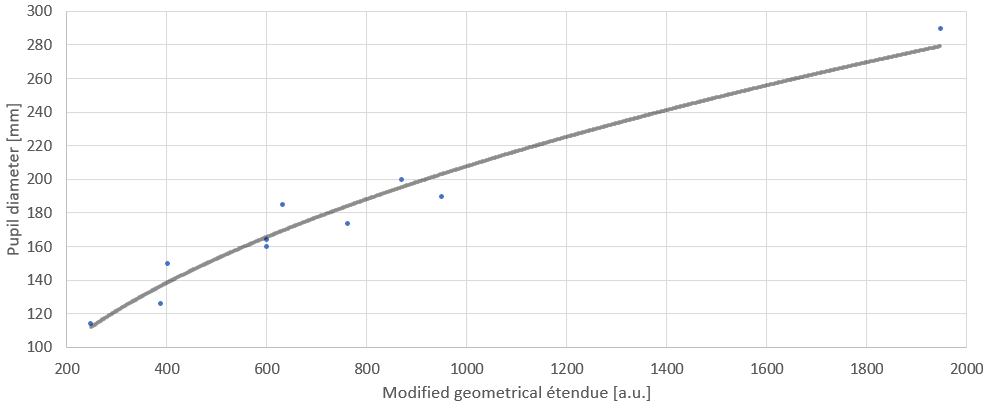}
\end{tabular}
\end{center}
\caption[SC]
{ \label{fig:D_pup_diop_app}
Fitting of the dioptric pupil diameters as a function of the modified geometrical étendue $x = A\Omega^{\delta}$.}
\end{figure}

The dioptric sample comprises: BlueMUSE\cite{BlueMUSE_P}, 4MOST\cite{4MOST_P}, WEAVE\cite{WEAVE_P}, DESI\cite{DESI_P}, SDSS\cite{SDSS_P}, MEGARA\cite{MEGARA_P}, WFOS\cite{WFOS_P}, Hector\cite{Hector_P}, WFMOS\cite{WFMOS_P}, IMACS\cite{IMACS_P}, and LLAMAS\cite{LLAMAS_P}. The power law of Eq.~\ref{eq:Dpup_mod} is fitted by maximising $R^2$ over the free exponent $\delta$, yielding $\alpha = 9.7163$~a.u., $\beta = 0.4434$, $\delta = 1.75$, $R^2 = 0.958$ (Fig.~\ref{fig:D_pup_diop_app}).

\subsubsection*{Catadioptric calibration}

\begin{figure} [H]
\begin{center}
\begin{tabular}{c}
\includegraphics[height=6cm]{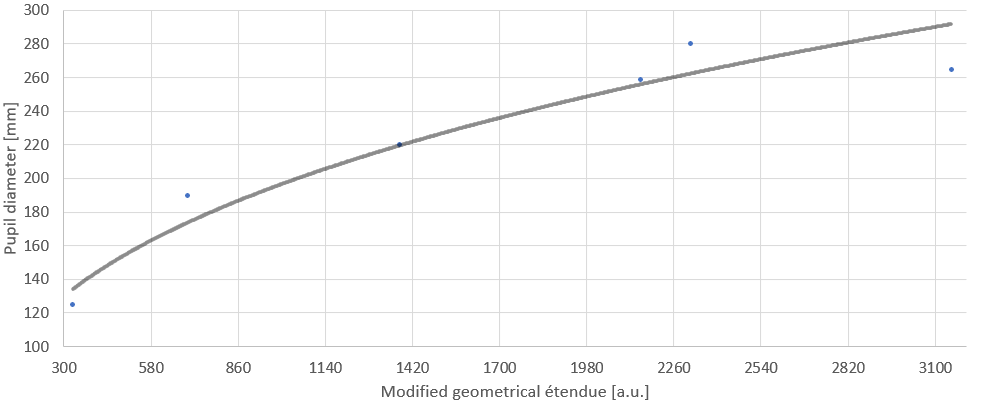}
\end{tabular}
\end{center}
\caption[SC]
{ \label{fig:D_pup_cata_app}
Fitting of the catadioptric pupil diameters as a function of the modified geometrical étendue $x = \gamma A\Omega^{\delta}$.}
\end{figure}

The catadioptric sample comprises: PFS\cite{PFS_P}, AAOmega\cite{AAOmega_P}, VIRUS\cite{VIRUS_P}, MOONS\cite{MOONS_P}, LAMOST\cite{LAMOST_P}, and FOBOS\cite{FOBOS_P}. The power law of Eq.~\ref{eq:Dpup_cata} is fitted analogously, yielding $\alpha = 18.329$~a.u., $\beta = 0.3436$, $\delta = 1.3$, $R^2 = 0.922$ (Fig.~\ref{fig:D_pup_cata_app}).

\subsection{Throughput and volume supporting derivations}
\label{app:obstruction}

This appendix provides two derivations that support the filtering steps of Sec.~\ref{sec:metrics}: the dependence of catadioptric obstruction on pupil diameter, and the parametric volume model used to estimate spectrograph envelope dimensions.

\subsubsection*{Additional obstruction as a function of pupil diameter}

\begin{figure} [H]
\begin{center}
\begin{tabular}{c}
\includegraphics[height=13cm]{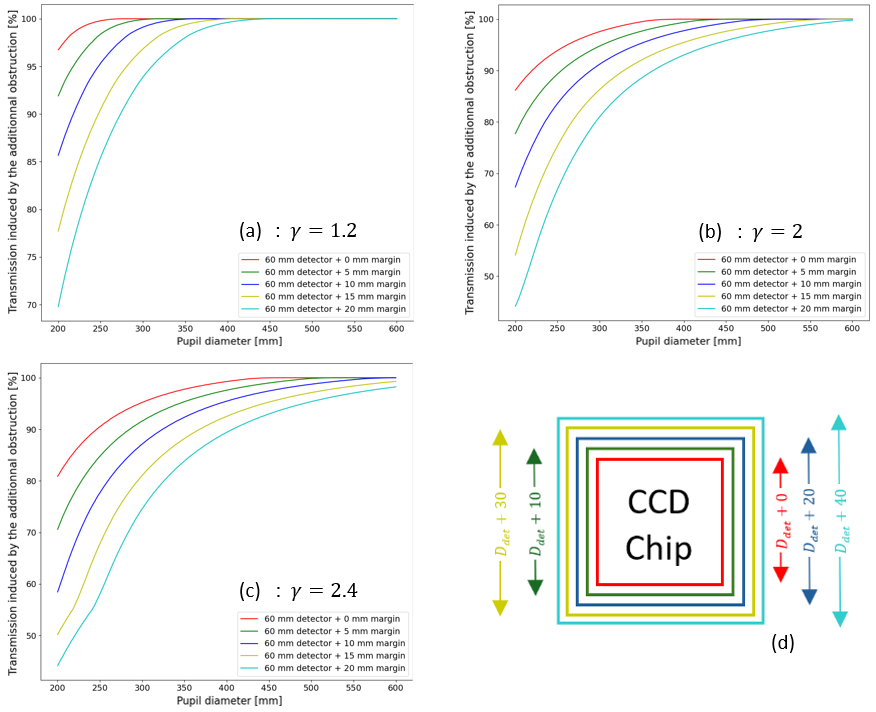}
\end{tabular}
\end{center}
\caption[SC]
{ \label{fig:obs_pup_app}
Additional obstruction induced by a square aperture containing a 60~mm detector with varying margins, as a function of pupil diameter. Each panel corresponds to a fixed anamorphism ratio: $\gamma = 1.2$ (a), $\gamma = 2$ (b), and $\gamma = 2.4$ (c). Panel (d) illustrates the detector geometry and margin definition.}
\end{figure}

Fig.~\ref{fig:obs_pup_app} shows the additional obstruction induced by a square aperture containing a 60~mm detector with varying margins, as a function of pupil diameter, for three representative anamorphism ratios. Two aspects are immediately apparent. First, a small change in the margin around the detector has a large effect on the additional obstruction, particularly at small pupil diameters. Second, the anamorphism ratio plays a decisive role: in an IFS, the fore-optics compress the pupil along one axis by a factor $\gamma$, which scales both the effective pupil size and the projected obstruction area accordingly. The difference in additional obstruction between $\gamma = 1.2$ and $\gamma = 2$ is already substantial, and would be even more pronounced compared to the unanamorphic case ($\gamma = 1$).

\subsubsection*{Volume toy model: parametric derivation and calibration}
\label{app:volume}

The model geometry is illustrated in Fig.~\ref{fig:paraxial_app}. From the paraxial layout, the collimator and camera envelope dimensions are:

\begin{figure} [H]
\begin{center}
\begin{tabular}{c}
\includegraphics[height=7cm]{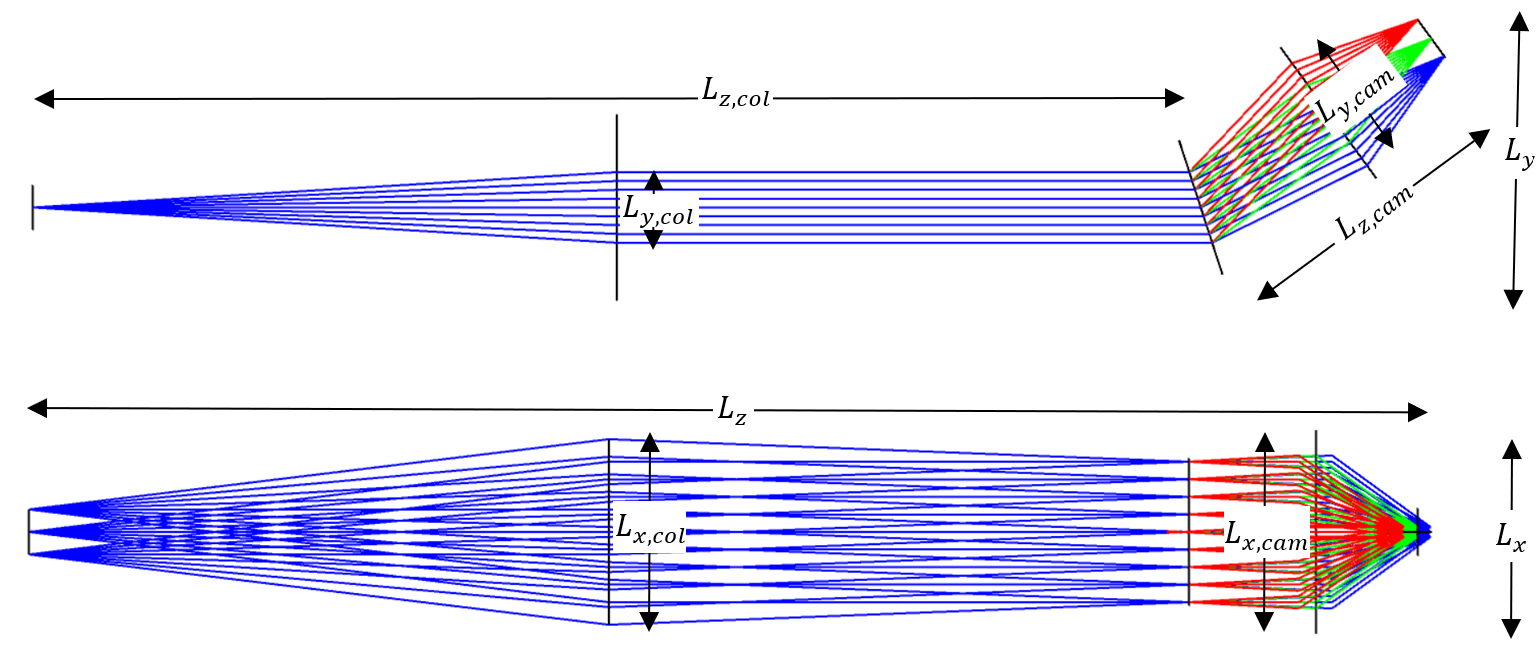}
\end{tabular}
\end{center}
\caption[SC]
{ \label{fig:paraxial_app}
Paraxial layout of a spectrograph used as the basis for the volume toy model.}
\end{figure}

\begin{align}
L_{x,\mathrm{col}} &= \left(D_\mathrm{det}\frac{f/_{col}}{f/_{cam}} + \tau_1 \min\!\left(1,\gamma\right)D_\mathrm{pupil}\right) m_\mathrm{opt}, 
\label{eq:Lxcol} \\[6pt]
L_{y,\mathrm{col}} &= \left(\min\!\left(1,\frac{1}{\gamma}\right) D_\mathrm{pupil}\right)m_\mathrm{opt}, 
\label{eq:Lycol} \\[6pt]
L_{x,\mathrm{cam}} &= L_{y,\mathrm{cam}} = \left(\min\!\left(1,\gamma\right) D_\mathrm{pupil} + \tau_2 f/_\mathrm{cam}\theta\right) m_\mathrm{opt}, 
\label{eq:Lxcam} \\[6pt]
L_{z,\mathrm{col}} &= \tau_3 D_\mathrm{pupil} f/_\mathrm{col}, 
\label{eq:Lzcol} \\[6pt]
L_{z,\mathrm{cam}} &= \tau_4 D_\mathrm{pupil} f/_\mathrm{cam}, 
\label{eq:Lzcam}
\end{align}

\noindent from which the overall envelope dimensions follow:

\begin{align}
L_x &= \max(L_{x,\mathrm{col}},\, L_{x,\mathrm{cam}}), \\[6pt]
L_y &= \frac{L_{y,\mathrm{col}}}{2} + \frac{L_{y,\mathrm{cam}}}{2}\cos(2\alpha) + L_{z,\mathrm{cam}} \sin(2\alpha), \\[6pt]
L_z &= L_{z,\mathrm{col}} + L_{z,\mathrm{cam}} \cos(2\alpha), \\[6pt]
V   &= \left(L_x L_y L_z\right) \, m_\mathrm{meca},
\label{eq:V}
\end{align}

\noindent where $\alpha$ is the grating angle with respect to the collimator and camera axes (Littrow configuration), $\gamma$ is the anamorphism ratio, $D_\mathrm{det}$ is the detector diameter, $\theta$ is the field angle at the pupil, and $f/_\mathrm{col}$, $f/_\mathrm{cam}$ are the f-numbers of the collimator and camera respectively. The margins $m_\mathrm{opt}$ and $m_\mathrm{meca}$ account respectively for the clear-to-mechanical aperture difference and for the additional volume required by the mechanics.

The four parameters $\tau_1$--$\tau_4$ are calibrated against MUSE, BlueMUSE, 4MOST, and MOONS, as summarised in Tab.~\ref{tab:tau_cal_app}. The model assumes a dioptric collimator, so $\tau_1$--$\tau_4$ are fully determined from MUSE and BlueMUSE, with 4MOST and MOONS providing partial calibration for the camera terms only. The same parameter set can be applied to catadioptric cameras by adapting $\tau_2$ and $\tau_4$ to that design type (obtained on MOONS). The adopted parameter values are $\tau_1 = 0.75$, $\tau_2 = 1$, $\tau_3 = 2$, $\tau_4 = 2.5$ for dioptric cameras, and $\tau_1 = 0.75$, $\tau_2 = 2.6$, $\tau_3 = 2$, $\tau_4 = 2.65$ for catadioptric cameras, with fixed parameters $\alpha = 10^\circ$, $f/_\mathrm{col} = 3.6$, $m_\mathrm{opt} = 1.1$, and $m_\mathrm{meca} = 2.6$.

\begin{table}[H]
\caption{Real and estimated dimensions for the calibration spectrographs, without optical or mechanical margins.}
\label{tab:tau_cal_app}
\begin{center}
\begin{tabular}{|>{\centering\arraybackslash}m{2.2cm}|
                 >{\centering\arraybackslash}m{1.6cm}|
                 >{\centering\arraybackslash}m{1.7cm}|
                 >{\centering\arraybackslash}m{1.7cm}|
                 >{\centering\arraybackslash}m{1.4cm}|
                 >{\centering\arraybackslash}m{1.7cm}|
                 >{\centering\arraybackslash}m{1.7cm}|
                 >{\centering\arraybackslash}m{1.4cm}|}
\hline
\multicolumn{2}{|c|}{\rule[-1ex]{0pt}{3.5ex}Instrument} 
    & \shortstack{$Lx_{col}/m_{opt}$\\{[mm]}} 
    & \shortstack{$Ly_{col}/m_{opt}$\\{[mm]}} 
    & \shortstack{$Lz_{col}$\\{[mm]}} 
    & \shortstack{$Lx_{cam}/m_{opt}$\\{[mm]}} 
    & \shortstack{$Ly_{cam}/m_{opt}$\\{[mm]}} 
    & \shortstack{$Lz_{cam}$\\{[mm]}} \\
\hline
\rule[-1ex]{0pt}{3.5ex} \multirow{2}{*}{MUSE} & Real & 194 & 65 & 856 & 154 & 102 & 372 \\
\cline{2-8}
\rule[-1ex]{0pt}{3.5ex} & Estimated & 193 & 49 & 865 & 150 & 101 & 375 \\
\hline
\rule[-1ex]{0pt}{3.5ex} \multirow{2}{*}{BlueMUSE} & Real & 222 & 90 & 920 & 156 & 123 & 467 \\
\cline{2-8}
\rule[-1ex]{0pt}{3.5ex} & Estimated & 222 & 87 & 908 & 148 & 119 & 471 \\
\hline
\rule[-1ex]{0pt}{3.5ex} \multirow{2}{*}{4MOST Blue} & Real & N/A & N/A & N/A & 224 & 224 & 610 \\
\cline{2-8}
\rule[-1ex]{0pt}{3.5ex} & Estimated & N/A & N/A & N/A & 226 & 226 & 612 \\
\hline
\rule[-1ex]{0pt}{3.5ex} \multirow{2}{*}{MOONS RI} & Real & N/A & N/A & N/A & 416 & 416 & 660 \\
\cline{2-8}
\rule[-1ex]{0pt}{3.5ex} & Estimated & N/A & N/A & N/A & 417 & 417 & 665 \\
\hline
\end{tabular}
\end{center}
\end{table}

The resulting volume estimates for all design options are given in Tab.~\ref{tab:volumes} below. The IFS room is modelled as a cylinder of diameter 22\,500~mm and height 6\,750~mm.

\begin{table}[H]
\caption{Estimated volume for the different design options, using the anamorphism ratios of Tab.~\ref{tab:A_list}. The IFS room is modelled as a cylinder of diameter 22\,500~mm and height 6\,750~mm.}
\label{tab:volumes}
\begin{center}
\begin{tabular}{|>{\centering\arraybackslash}m{2.0cm}|
                 >{\centering\arraybackslash}m{1.9cm}|
                 >{\centering\arraybackslash}m{1.8cm}|
                 >{\centering\arraybackslash}m{2.2cm}|
                 >{\centering\arraybackslash}m{2.2cm}|
                 >{\centering\arraybackslash}m{2.0cm}|
                 >{\centering\arraybackslash}m{1.5cm}|}
\hline
\rule[-1ex]{0pt}{3.5ex} Design option & Camera design & Sampling & Number of spectrographs & Volume per spectrograph (2 arms) [$\mathrm{m^3}$] & Total volume [$\mathrm{m^3}$] & \% of IFS room \\
\hline
\rule[-1ex]{0pt}{3.5ex} \multirow{4}{*}{4k-15µm} & \multirow{2}{*}{Dioptric} & 0.25"$\times$0.25" & 144 & 3.3 & 472 & 17.6 \\
\cline{3-7}
\rule[-1ex]{0pt}{3.5ex}  &  & 0.3"$\times$0.2" & 150 & 1.6 & 239 & 8.9 \\
\cline{2-7}
\rule[-1ex]{0pt}{3.5ex}  & \multirow{2}{*}{Catadioptric} & 0.25"$\times$0.25" & 144 & 4.6 & 659 & 24.5 \\
\cline{3-7}
\rule[-1ex]{0pt}{3.5ex}  &  & 0.3"$\times$0.2" & 150 & 2.2 & 334 & 12.5 \\
\hline
\rule[-1ex]{0pt}{3.5ex} \multirow{4}{*}{6k-10µm} & \multirow{2}{*}{Dioptric} & 0.25"$\times$0.25" & 96 & 14.0 & 1342 & 50.0 \\
\cline{3-7}
\rule[-1ex]{0pt}{3.5ex}  &  & 0.3"$\times$0.2" & 100 & 6.7 & 667 & 24.9 \\
\cline{2-7}
\rule[-1ex]{0pt}{3.5ex}  & \multirow{2}{*}{Catadioptric} & 0.25"$\times$0.25" & 96 & 11.6 & 1115 & 41.5 \\
\cline{3-7}
\rule[-1ex]{0pt}{3.5ex}  &  & 0.3"$\times$0.2" & 100 & 5.7 & 566 & 21.1 \\
\hline
\rule[-1ex]{0pt}{3.5ex} \multirow{4}{*}{6k-15µm} & \multirow{2}{*}{Dioptric} & 0.25"$\times$0.25" & 96 & 9.6 & 917 & 34.2 \\
\cline{3-7}
\rule[-1ex]{0pt}{3.5ex}  &  & 0.3"$\times$0.2" & 100 & 4.5 & 449 & 16.7 \\
\cline{2-7}
\rule[-1ex]{0pt}{3.5ex}  & \multirow{2}{*}{Catadioptric} & 0.25"$\times$0.25" & 96 & 13.0 & 1245 & 46.4 \\
\cline{3-7}
\rule[-1ex]{0pt}{3.5ex}  &  & 0.3"$\times$0.2" & 100 & 6.3 & 626 & 23.3 \\
\hline
\rule[-2ex]{0pt}{3.5ex} \multirow{4}{*}{\shortstack{6k-10µm\\(2x1 bin.)}} & \multirow{2}{*}{Dioptric} & 0.25"$\times$0.25" & 192 & 1.3 & 247 & 9.2 \\
\cline{3-7}
\rule[-1ex]{0pt}{3.5ex}  &  & 0.3"$\times$0.2" & 200 & 0.9 & 176 & 6.6 \\
\cline{2-7}
\rule[-1ex]{0pt}{3.5ex}  & \multirow{2}{*}{Catadioptric} & 0.25"$\times$0.25" & 192 & 1.9 & 356 & 13.3 \\
\cline{3-7}
\rule[-1ex]{0pt}{3.5ex}  &  & 0.3"$\times$0.2" & 200 & 0.9 & 186 & 6.9 \\
\hline
\rule[-2ex]{0pt}{3.5ex} \multirow{4}{*}{\shortstack{8k-10µm\\(2x1 bin.)}} & \multirow{2}{*}{Dioptric} & 0.25"$\times$0.25" & 144 & 2.8 & 410 & 15.3 \\
\cline{3-7}
\rule[-1ex]{0pt}{3.5ex}  &  & 0.3"$\times$0.2" & 150 & 2.0 & 293 & 10.9 \\
\cline{2-7}
\rule[-1ex]{0pt}{3.5ex}  & \multirow{2}{*}{Catadioptric} & 0.25"$\times$0.25" & 144 & 3.7 & 526 & 19.6 \\
\cline{3-7}
\rule[-1ex]{0pt}{3.5ex}  &  & 0.3"$\times$0.2" & 150 & 1.8 & 264 & 9.8 \\
\hline
\rule[-2ex]{0pt}{3.5ex} \multirow{4}{*}{\shortstack{9k-10µm\\(2x1 bin.)}} & \multirow{2}{*}{Dioptric} & 0.25"$\times$0.25" & 128 & 4.1 & 519 & 19.3 \\
\cline{3-7}
\rule[-1ex]{0pt}{3.5ex}  &  & 0.3"$\times$0.2" & 134 & 2.8 & 373 & 13.9 \\
\cline{2-7}
\rule[-1ex]{0pt}{3.5ex}  & \multirow{2}{*}{Catadioptric} & 0.25"$\times$0.25" & 128 & 4.7 & 598 & 22.3 \\
\cline{3-7}
\rule[-1ex]{0pt}{3.5ex}  &  & 0.3"$\times$0.2" & 134 & 2.2 & 295 & 11.0 \\
\hline
\rule[-2ex]{0pt}{3.5ex} \multirow{4}{*}{\shortstack{9k-10µm\\(3x1 bin.)}} & \multirow{2}{*}{Dioptric} & 0.25"$\times$0.25" & 192 & 2.1 & 398 & 14.8 \\
\cline{3-7}
\rule[-1ex]{0pt}{3.5ex}  &  & 0.3"$\times$0.2" & 200 & 1.3 & 259 & 9.6 \\
\cline{2-7}
\rule[-1ex]{0pt}{3.5ex}  & \multirow{2}{*}{Catadioptric} & 0.25"$\times$0.25" & 192 & 2.3 & 448 & 16.7 \\
\cline{3-7}
\rule[-1ex]{0pt}{3.5ex}  &  & 0.3"$\times$0.2" & 200 & 2.2 & 439 & 16.4 \\
\hline
\rule[-2ex]{0pt}{3.5ex} \multirow{4}{*}{\shortstack{6k-15µm\\(2x1 bin.)}} & \multirow{2}{*}{Dioptric} & 0.25"$\times$0.25" & 192 & 1.0 & 184 & 6.9 \\
\cline{3-7}
\rule[-1ex]{0pt}{3.5ex}  &  & 0.3"$\times$0.2" & 200 & 0.7 & 135 & 5.0 \\
\cline{2-7}
\rule[-1ex]{0pt}{3.5ex}  & \multirow{2}{*}{Catadioptric} & 0.25"$\times$0.25" & 192 & 2.3 & 441 & 16.4 \\
\cline{3-7}
\rule[-1ex]{0pt}{3.5ex}  &  & 0.3"$\times$0.2" & 200 & 1.2 & 231 & 8.6 \\
\hline
\end{tabular}
\end{center}
\end{table}